\documentclass[a4paper,11pt]{article}

\usepackage[numbers,sort&compress]{natbib}
\usepackage{jcappub} 

\usepackage[T1]{fontenc} 

\usepackage{latexsym}
\usepackage{graphicx}
\usepackage{amsmath}
\usepackage{amssymb}
\usepackage{amsfonts}
\usepackage{times}
\usepackage{xspace} 

\usepackage{subfigure}		
\usepackage[normalem]{ulem} 

\usepackage{hyperref} 

\title{\boldmath Self-tuning kinetic gravity braiding: Cosmological dynamics, shift symmetry, and the tadpole}

\author{Reginald Christian Bernardo}
\affiliation{National Institute of Physics, University of the Philippines, Diliman, Quezon City 1101, Philippines}

\emailAdd{rbernardo@nip.upd.edu.ph}

\abstract{We study the self-tuning subclass of kinetic gravity braiding and obtain robust predictions on self-tuning and dynamics in the tadpole-free shift symmetric sector of the theory. In particular, we show inevitability of cosmic acceleration, prove the dynamical stability of this late-time asymptotic state, and derive ghost and gradient stability constraints on the self-tuning vacuum. We discuss the results concretely in the context of generalized cubic covariant Galileon theory and an exponential kinetic gravity braiding.
}

\begin{document}
\maketitle
\flushbottom

\section{Introduction}
\label{sec:intro}

The $\Lambda$CDM model is considered to be the best theory of the observable Universe. However, it leaves the existence of optically invisible cosmic fluids in question and suffers from the titanic disparity between the observed late-time cosmic acceleration and vacuum energy. It is therefore worthwhile to investigate modified theories of gravity that for Solar system scales reduce to general relativity (GR) -- the gravitational cornerstone of the $\Lambda$CDM model -- but with a cosmologically-active dark degree of freedom. In the context of the cosmological constant problem, it is also interesting to look for the very special sectors of these modified theories that are accompanied by self-adjusting or \textit{self-tuning} fields which eat up the dependence of  cosmic acceleration on vacuum energy. 

Kinetic gravity braiding (KGB) belongs to the remarkable class of scalar-tensor theories with second-order field equations which feature self-tuning, screening, and scaling mechanisms \cite{st_galileon_inflation_deffayet, st_galileon_inflation_kobayashi, st_horndeski_galileons_charmousis, st_horndeski_cubic_appleby, st_horndeski_cosmology_emond2018, st_horndeski_variations_appleby, st_fugue_appleby, screening_mcmanus, kgb_vainshtein_anson, st_horndeski_scaling_solutions_albuquerque, scaling_kgb_frusciante}. Its phenomenological power was showcased when it surpassed the very stringent constraint on nearly-luminal gravitational wave propagation \cite{gw_170817_ligo, lombriser2016breaking, st_horndeski_cosmology_sakstein, st_horndeski_cosmology_baker, dark_energy_creminelli, dark_energy_ezquiaga} and it remains to be a viable dark energy theory even with cosmological constraints \cite{st_horndeski_galileon_barreira_1, st_horndeski_galileon_barreira_2, st_horndeski_renk,  st_horndeski_constraint_mancini2019, horndeski_constraint_noller2018, st_horndeski_constraint_komatsu2019, st_horndeski_galileon_peirone}. Recent results even show that the Galileon ghost condensate \cite{peirone2019cosmological} and the generalized cubic covariant Galileon \cite{gccg_frusciante} theories of the tadpole-free shift symmetric sector of KGB can be as compatible with cosmological observations as the $\Lambda$CDM model. Interestingly, the same sector also houses hairy black holes and could be potentially rich with observational signatures from the strong gravity regime \cite{bernardo2019hair, stealth_bernardo, self_tuning_bh_emond}. 

Remakably, the dark energy field itself can be endowed with a self-tuning mechanism in order to both produce cosmic acceleration and protect the spacetime from the influence of a bare cosmological constant. This was implemented in the Fab Four theories by making the field equations trivally satisfied on Minkowski vacuum and evading Weinberg's no-go theorem through a time-dependent scalar field \cite{st_horndeski_galileons_charmousis}. However, it turns out that this implementation of self-tuning screens all kinds of energy and as a consequence outputs theories with a questionable viability of matter and radiation eras. A more recent alternative, and arguably an improvement, to the Fab Four type of self-tuning was realized in the form of well-tempering \cite{st_horndeski_cubic_appleby}. In this approach, the Hubble and scalar field equations are compelled to be degenerate on a de Sitter state, and so by construction screens only vacuum energy and evades Weinberg's no-go theorem by breaking time-translation invariance \cite{st_horndeski_cubic_appleby}. This was introduced in the context of tadpole-driven shift symmetric KGB in Ref. \cite{st_horndeski_cubic_appleby} and later extended to the broader shift symmetry breaking sector including a conformal coupling with the Ricci scalar in Refs. \cite{st_horndeski_cosmology_emond2018, st_horndeski_variations_appleby}. An interesting recent work also demonstrates how well-tempering can be used to deliver only a Minkowski vacuum in scalar-tensor theories \cite{st_fugue_appleby}.

In general, both the Fab Four type (later referred to as the trivial scalar approach) and well-tempering implementations of self-tuning in a de Sitter background can be accommodated in KGB and broader scalar-tensor theories (figure \ref{fig:kgb_self_tuning}). The exception to this is the tadpole-free shift symmetric sector which apparently will not admit a well-tempering scalar field. This has been recognized in Ref. \cite{st_horndeski_cubic_appleby} which devoted its Appendix B to tease its solution. In this work, we put this realization in a much stronger \textit{no-tempering} theorem (Sec. \ref{subsec:well_tempered_kgb}) and set out to fully explore the self-tuning branch of tadpole-free shift symmetric KGB.

The rest of this paper proceeds as follows. We start by introducing KGB and the relevant astrophysical and cosmological constraints (Sec. \ref{sec:kgb}). We then focus on shift symmetric KGB and explain in detail both well-tempering (Sec. \ref{subsec:well_tempered_kgb}) and trivial scalar (Sec. \ref{subsec:self_tuned_kgb}) approaches to self-tuning. Owing to the no-tempering theorem derived in Section \ref{subsec:well_tempered_kgb}, we restrict the rest of the discussion to the self-tuning theory resulting from trivial scalar approach. We prove that its de Sitter state is a stable fixed point (Sec. \ref{subsec:stability_ds}), establish an important connection between the de Sitter vacuum and the vanishing shift current hypersurface (Sec. \ref{subsec:self_tuning_tailoring}), and provide ghost and gradient stability constraints on the theory (Sec. \ref{subsec:ghost_laplace_stability}). In Secs. \ref{sec:stgccg} and \ref{sec:cc_vs_gradient}, we present numerical experiments on the generalized cubic covariant Galileon theory and an exponential KGB with a mixture of cosmic fluids, confirming the analytical results and addressing the viability of tuning away an arbitrarily large bare vacuum energy. We end with a discussion of key issues and future work (Secs. \ref{sec:tempered_trivial_scalar} and \ref{sec:outlook}).

\textit{Conventions.} We work with the mostly plus metric signature, $\left( -, + ,+ ,+ \right)$ and units with $c = M_P^2 = 1$ where $M_P^2 = 1/8 \pi G$, $c$ is the speed of light, and $G$ is Newton's gravitation constant. We use lowercase latin letters for spacetime indices and, for brevity, write down $\xi_a =\nabla_a \xi$, $\xi^a = \nabla^a \xi$, and $\xi_{ab} = \nabla_b \nabla_a \xi$ for any scalar function $\xi$, where $\nabla_a$ is the covariant derivative. The calculations performed in this paper have been performed both by hand and using Mathematica, exploiting the xAct \cite{xact}, xPert \cite{xpert}, xCoba \cite{xcoba}, and xPand \cite{xpand} packages for abstract tensor calculus and for deriving the field equations and its perturbations in the FRW metric. The reader interested in all the mathematical details of this paper is encouraged to download the supporting notebooks in the \href{https://reggiebernardo.weebly.com/notebooks.html}{author's website}.

\section{Kinetic gravity braiding}
\label{sec:kgb}

Kinetic gravity braiding (KGB) is given by the gravitational action \cite{st_galileon_inflation_deffayet, st_galileon_inflation_kobayashi}
\begin{equation}
\label{eq:kgb}
S_g = \int d^4 x \sqrt{-g} \left[ \kappa R + K \left( \phi, X \right) - G \left( \phi, X \right) \Box \phi \right]
\end{equation}
where $g_{ab}$ is the metric, $R$ is the Ricci scalar, $\kappa = M_P^2 / 2$, $\phi$ is the scalar field, $X$ is the scalar field's kinetic density given by $X = - g^{ab} \phi_a \phi_b / 2$, and $K$ and $G$ are the $k$-essence and braiding potentials, respectively, both of which are arbitrary functions of their arguments \footnote{``Braiding" refers to the mixing of scalar ($\psi$) and tensor ($h$) modes present in the term $G \Box \phi \sim G \, \partial h \, \partial \psi + O\left( h^2, \psi^2 \right)$ in the second order action which entangles the scalar and tensor modes \cite{st_galileon_inflation_deffayet}.}. This theory is a large swath of Horndeski theory with gravitational waves that propagate at the speed of light \footnote{A more general theory with only luminally-propagating tensor modes and second-order field equations would involve a conformal coupling, $S \sim \int d^4 x \sqrt{-g} f\left( \phi \right)R$ between the scalar field and the Ricci scalar in the action. This broader sector includes Brans-Dicke theory and $f\left(R\right)$ gravity. There is also a very special class of quartic and quintic Horndeski theories with luminally-propagating tensor modes \cite{st_horndeski_copeland2018}.} \cite{st_horndeski_seminal, st_galileon_inflation_kobayashi_2, st_dark_energy_tsujikawa, horndeski_review_kobayashi2019}. Some known subsets of KGB include quintessence $\left( K = X - V\left( \phi \right), G = \text{constant} \right)$ \cite{quintessence_building_tsamis, quintessence_building_saini}, $k$-essence $\left( K = K\left( \phi , X \right) , G = \text{constant} \right)$ \cite{kessence_seminal_armendariz, kessence_seminal_armendariz2}, cubic Galileon theory $\left( K \sim X, G \sim X \right)$ \cite{st_galileon_inflation_burrage}, generalized cubic covariant Galileon $\left( K \sim X^{p_2}, G \sim X^{p_3} \right)$ \cite{st_horndeski_cosmology_de_felice}, and the Galileon ghost condensate $\left( K \sim X + X^2, G \sim X \right)$ \cite{ggc_dark_energy_kase}. The shift symmetric sector of the theory, invariant under the shift transformation, $\phi \rightarrow \phi + k$ for constant $k$, is a special class for potentially controlling radiative corrections. The cubic Galileon theory, generalized cubic covariant Galileon, and Galileon ghost condensate belong with KGB's shift symmetric class.

Several existing constraints make KGB a phenomenologically powerful theory. First of all, screening mechanisms exist in KGB to suppress the dark energy field at Solar system scales \cite{screening_mcmanus, kgb_vainshtein_anson}. Second, among few others, KGB bypasses the very stringent constraint on the gravitational wave speed \footnote{This constraint must be carefully interpreted from an effective field theory stand point \cite{dark_energy_derham}.} \cite{lombriser2016breaking, st_horndeski_cosmology_sakstein, st_horndeski_cosmology_baker, dark_energy_creminelli, dark_energy_ezquiaga}. Third, cosmological constraints continue to argue for and against selected KGB sectors, for instance, disfavoring cubic Galileon theory while controversially preferring Galileon ghost condensate over the $\Lambda$CDM theory \cite{st_horndeski_galileon_barreira_1, st_horndeski_galileon_barreira_2, st_horndeski_renk, st_horndeski_galileon_peirone, peirone2019cosmological, st_horndeski_constraint_mancini2019, horndeski_constraint_noller2018, st_horndeski_constraint_komatsu2019}. It has also been recently shown that generalized cubic covaraint Galileon is compatible with cosmological data \cite{gccg_frusciante}. The existing and future astrophysical and cosmological surveys are expected to tighten existing constraints in KGB and its broader extensions. Notably, the final theory can be one of KGB even if some observational data continue to support GR \cite{no_slip_linder, no_slip_cmb_brush, no_run_linder}.

For the rest of this work, we consider a spatially-flat Friedmann-Robertson-Walker (FRW) metric
\begin{equation}
\label{eq:frw_metric}
ds^2 = -dt^2 + a\left( t \right)^2 d \vec{ x }^2 
\end{equation}
and a comoving scalar field
\begin{equation}
\label{eq:comoving_scalar}
\phi = \phi \left( t \right) .
\end{equation}
The field equations can be obtained by varying the action \eqref{eq:kgb} with respect to the metric $g_{ab}$ and the scalar field $\phi$. Assuming that the matter sector is given by a perfect fluid, then the field equations can be written as
\begin{equation}
\label{eq:friedmann_eq}
3 H^2 = \rho - K + \dot{ \phi }^2 K_X + 3 H \dot{\phi}^3 G_X - \dot{\phi}^2 G_\phi ,
\end{equation}
\begin{equation}
\label{eq:hubble_eq}
2 \dot{H} = - \left( \rho + P \right) - \dot{\phi}^2 K_X - 3 H \dot{\phi}^3 G_X + \dot{\phi}^2 \ddot{\phi} G_X + 2 \dot{\phi}^2 G_\phi ,
\end{equation}
and
\begin{equation}
\label{eq:sfe}
\begin{split}
\ddot{ \phi } K_X & + 3 H \dot{ \phi } K_X - K_\phi + \dot{\phi}^2 \ddot{\phi} K_{XX} + \dot{\phi}^2 K_{\phi X} \\
& + \ddot{ \phi } \left( 6 H \dot{\phi} G_X + 3 H \dot{\phi}^3 G_{XX} - \dot{\phi}^2 G_{\phi X} - 2 G_\phi \right) \\
& + 9 H^2 \dot{\phi}^2 G_X - 6 H \dot{\phi} G_\phi + 3 H \dot{\phi}^3 G_{\phi X} + 3 \dot{\phi}^2 \dot{H} G_X - \dot{\phi}^2 G_{\phi \phi} = 0
\end{split}
\end{equation}
where $\rho$ and $P$ are the energy density and pressure of the perfect fluid, a dot over a variable refers to differentiation with respect to the comoving time $t$ and subscripts of $\phi$ and $X$ in the potentials refer to differentiation with respect to $\phi$ and $X$ in their arguments, e.g., $K_X = \partial K /\partial X$ and $K_{\phi X} = \partial^2 K / \partial \phi \partial X $. We refer to Eqs. \eqref{eq:friedmann_eq}, \eqref{eq:hubble_eq}, and \eqref{eq:sfe} as the Friedmann constraint, Hubble equation, and scalar field equation, respectively. In addition, the fluid is assumed to fall on the geodesics of the spacetime, on which it obeys the first law of thermodynamics which can be written as
\begin{equation}
\label{eq:geodesic_eq}
\dot{\rho} + 3 H \left( \rho + P \right) = 0 .
\end{equation}
Noninteracting fluids, $\rho_i$, would simply fall on their own independent trajectories. The system of equations given by Eqs. \eqref{eq:friedmann_eq}, \eqref{eq:hubble_eq}, \eqref{eq:sfe}, and \eqref{eq:geodesic_eq} can be closed by providing the equation of state $P \left( \rho \right)$ which amounts to specifying the types of fluid present in the universe, e.g., for a barotropic perfect fluid $P / \rho= w$ where $w = 0, 1/3, -1$ correspond to a nonrelativistic matter-, radiation-, and cosmogical constant-filled universes, respectively.

\section{Self-tuning kinetic gravity braiding}
\label{sec:self_tuned_kgb}

In this section, we draw the main results of this paper. We first discuss tempered self-tuning in tadpole-driven shift symmetric KGB (Sec. \ref{subsec:well_tempered_kgb}). This sets the stage for trivial scalar type self-tuning which takes the spotlight in the study of KGB's tadpole-free shift symmetric limit (Sec. \ref{subsec:self_tuned_kgb}). We prove the dynamical stability of the de Sitter vacuum (Sec. \ref{subsec:stability_ds}), establish an important connection between self-tuning and the vanishing scalar current hypersurface (Sec. \ref{subsec:self_tuning_tailoring}), and derive ghost and gradient stability constraints on the de Sitter vacuum (Sec. \ref{subsec:ghost_laplace_stability}).

\subsection{Well-tempered kinetic gravity braiding}
\label{subsec:well_tempered_kgb}

Tempered self-tuning, or \textit{well-tempering}, is a dynamical cancellation of vacuum energy introduced in Ref. \cite{st_horndeski_cubic_appleby} and applied to the tadpole-driven shift symmetric KGB given by
\begin{eqnarray}
\label{eq:k_tadpole} K\left( \phi, X \right) &=& l \phi + A \left( X \right) \\
\label{eq:g_tadpole} G\left( \phi, X \right) &=& B\left( X \right)
\end{eqnarray}
where $l$ is a constant and $A$ and $B$ are arbitrary functions of the scalar field's kinetic density $X$. The subset of KGB accommodating this mechanism can be found by searching for the theory space in which the Hubble and scalar field equations (Eqs. \eqref{eq:hubble_eq} and \eqref{eq:sfe}) reduce to the same equation on the de Sitter state $H\left( t \right) = h$ or ``on-shell'' \footnote{
The degeneracy between Eqs. \eqref{eq:hubble_eq} and \eqref{eq:sfe} is a natural choice because the Hubble equation depends only on the combination $\rho + P$ which trivially vanishes for vacuum energy, i.e., $\rho_\Lambda + P_\Lambda = 0$.
}. This can be broken into two steps: (1) taking the ``on-shell" limit of the field equations and (2) solving for a system of coupled differential equations for the potentials $A$ and $B$ determining the tempered branch. 

Using Eqs. \eqref{eq:k_tadpole} and \eqref{eq:g_tadpole} the field equations (Eqs. \eqref{eq:friedmann_eq}, \eqref{eq:hubble_eq}, and \eqref{eq:sfe}) reduce to
\begin{equation}
\label{eq:friedmann_eq_tadpole}
3 H^2 = \rho - l\phi - A + \dot{ \phi }^2 A_X + 3 H \dot{\phi}^3 B_X ,
\end{equation}
\begin{equation}
\label{eq:hubble_eq_tadpole}
2 \dot{H} = - \left( \rho + P \right) - \dot{\phi}^2 A_X - 3 H \dot{\phi}^3 B_X + \dot{\phi}^2 \ddot{\phi} B_X ,
\end{equation}
and
\begin{equation}
\label{eq:sfe_tadpole}
\begin{split}
\ddot{ \phi } A_X + 3 H \dot{ \phi } A_X - l + \dot{\phi}^2 \ddot{\phi} A_{XX} + \ddot{ \phi } H \left( 6 \dot{\phi} B_X + 3 \dot{\phi}^3 B_{XX} \right) + 9 H^2 \dot{\phi}^2 B_X + 3 \dot{\phi}^2 \dot{H} B_X = 0 ,
\end{split}
\end{equation}
respectively. [Step (1)] Taking the de Sitter limit, $H \left( t \right) = h$, the above system of equations become
\begin{equation}
\label{eq:friedmann_eq_tadpole_ds}
3 h^2 = \rho - l\phi - A + \dot{ \phi }^2 A_X + 3 h \dot{\phi}^3 B_X ,
\end{equation}
\begin{equation}
\label{eq:hubble_eq_tadpole_ds}
0 = - \left( \rho + P \right) - \dot{\phi}^2 A_X - 3 h \dot{\phi}^3 B_X + \dot{\phi}^2 \ddot{\phi} B_X ,
\end{equation}
and
\begin{equation}
\label{eq:sfe_tadpole_ds}
\begin{split}
\ddot{ \phi } A_X + 3 h \dot{ \phi } A_X - l + \dot{\phi}^2 \ddot{\phi} A_{XX} + \ddot{ \phi } h \left( 6 \dot{\phi} B_X + 3 \dot{\phi}^3 B_{XX} \right) + 9 h^2 \dot{\phi}^2 B_X = 0 .
\end{split}
\end{equation}
Then, for vacuum energy, $P_\Lambda = - \rho_\Lambda$, we express the Hubble and scalar field equations as
\begin{equation}
\dot{\phi}^2 A_X + 3 h \dot{\phi}^3 B_X - \ddot{\phi} \left( \dot{\phi}^2 B_X \right) = 0
\end{equation}
and
\begin{equation}
\begin{split}
- 3 h \dot{ \phi } A_X - 9 h^2 \dot{\phi}^2 B_X + l - \ddot{\phi} \left(  A_X + \dot{\phi}^2 A_{XX}  + 6 h \dot{\phi} B_X + 3h \dot{\phi}^3 B_{XX} \right) = 0 .
\end{split}
\end{equation}
[Step (2)] To obtain the common theory space shared by both equations on-shell,
we make the terms attached to $\ddot{\phi}$ proportional to each other. This leads to
\begin{eqnarray}
f\left( \dot{\phi} \right) \dot{\phi}^2 B_X &=& A_X + \dot{\phi}^2 A_{XX}  + 6 h \dot{\phi} B_X + 3h \dot{\phi}^3 B_{XX}  \\
- f\left( \dot{\phi} \right) \left( \dot{\phi}^2 A_X + 3 h \dot{\phi}^3 B_X \right) &=& 3 h \dot{ \phi } A_X + 9 h^2 \dot{\phi}^2 B_X - l 
\end{eqnarray}
where $f \left( x \right)$ is an arbitrary function that parametrizes the common theory space. Expressing this in terms of the kinetic density $X = \dot{\phi}^2/2$, we obtain a coupled, linear, first-order differential equations for $A_X$ and $B_X$:
\begin{equation}
\label{eq:condition_1}
q\left(X\right) \sqrt{2X} B_X = \dfrac{d}{dX} \left[ 2 X \left( A_X + 3 h \sqrt{2X} B_X \right) \right] - \left( A_X + 3 h \sqrt{2X} B_X \right)
\end{equation}
and
\begin{equation}
\label{eq:condition_2}
- q\left( X \right) \sqrt{2X} \left( A_X + 3 h \sqrt{2 X} B_X \right) = 3 h \sqrt{2X} \left( A_X + 3 h \sqrt{2 X} B_X \right) - l 
\end{equation}
where $q\left( X \right) = \sqrt{2X} f\left( \sqrt{2X} \right)$ \footnote{These are Eqs. (24) and (25) in Ref. \cite{st_horndeski_cubic_appleby}.}. The solution to this for $l \neq 0$ is given by \footnote{It is easy to obtain this after noticing the system's explicit dependence on $\left( A_X + 3h \sqrt{2X} B_X \right)$.}
\begin{eqnarray}
\label{eq:A_wt} A_X &=& \dfrac{ l \left( 3 h q\left(X\right) + q\left(X\right)^2 + 6 h X q'\left(X\right) \right) }{ \sqrt{2X} q\left(X\right) \left(3h + q\left(X\right)\right)^2 } \\
\label{eq:B_wt} B_X &=& - \dfrac{ l q'\left(X\right) }{ q\left(X\right) \left(3h + q\left(X\right)\right)^2 } .
\end{eqnarray}
Substituting this back into the on-shell Hubble and scalar field equations, and noting that $\dot{X} = \dot{\phi}\ddot{\phi}$ and $\dot{q} = q'\left(X\right) \dot{X}$, we obtain the Riccati equation
\begin{equation}
\label{eq:riccati}
\dot{y} \left( t \right) + y \left( t \right) \left( 3 h + y \left( t \right) \right) = 0
\end{equation}
with the exact solution
\begin{equation}
\label{eq:riccati_sol}
q \left( \dfrac{ \dot{ \phi }^2 }{2} \right) = \dfrac{ 3 h }{ \exp\left( 3h \left( t - T \right) \right) -1 }
\end{equation}
where $y\left( t \right) = q\left( \dot{\phi}^2/2 \right)$ and $T$ is an integration constant. It is important to point out that the Friedmann constraint becomes
\begin{equation}
\label{eq:friedmann_eq_wt}
3 h^2 = \rho_\Lambda - A\left( \dot{ \phi }^2 / 2 \right) - l \phi + \dfrac{ l \dot{\phi} }{ 3 h + q\left( \dot{ \phi }^2 / 2 \right) }
\end{equation}
and reflects the dynamical cancellation of vacuum energy through its nontrivial time-dependence. By differentiating this equation with respect to the comoving time $t$, we obtain the same Riccati equation (Eq. \eqref{eq:riccati}). Thus, for the well-tempered theory, parametrized by Eq. \eqref{eq:A_wt} and \eqref{eq:B_wt}, all field equations (Eqs. \eqref{eq:friedmann_eq_tadpole}, \eqref{eq:hubble_eq_tadpole}), and \eqref{eq:sfe_tadpole}) reduce to the same Riccati equation (Eq. \eqref{eq:riccati}) on-shell and the exact solution $\left( H\left(t\right), \phi\left(t\right) \right)$ is given by $H\left( t \right) = h$ and Eq. \eqref{eq:riccati_sol}.

This tempered regime has been extended to the larger shift symmetry breaking sector of KGB including a conformal coupling with the Ricci scalar in the gravitational action \cite{st_horndeski_cosmology_emond2018, st_horndeski_variations_appleby}. However, it turns out that tempering cannot be achieved in the tadpole-free limit. We prove this with the following short theorem.

\subsubsection*{\textit{No-tempering in the tadpole-free shift symmetric KGB}}

The parametrization of tempered KGB given by Eqs. \eqref{eq:A_wt} and \eqref{eq:B_wt} naively suggests that $A_X = 0$ and $B_X = 0$ (trivial theory) in the tadpole-free limit. A careful look back at Eq. \eqref{eq:condition_2} instead shows that $l = 0$ implies $q \left( X \right) = -3 h$ and so the breakdown of Eqs. \eqref{eq:A_wt} and \eqref{eq:B_wt} can be recognized to be merely an artifact of the parametrization.

To get to the \textit{possible} tadpole-free tempering subclass of theory \eqref{eq:kgb}, we go back to Eqs. \eqref{eq:condition_1} and \eqref{eq:condition_2} which reduces to the single equation
\begin{equation}
\label{eq:condition_l0}
-3 h \sqrt{2X} B_X = \dfrac{d}{dX} \left[ 2 X \left( A_X + 3 h \sqrt{2X} B_X \right) \right] - \left( A_X + 3 h \sqrt{2X} B_X \right) 
\end{equation}
in the limit $l \rightarrow 0$. This has the exact solution
\begin{equation}
\label{eq:almost_temp_theory}
B \left[ A \right] = - \dfrac{ 2 \alpha }{\sqrt{X}} + \beta - \dfrac{A}{3 \sqrt{2} h \sqrt{X}} 
\end{equation}
where $\alpha$ and $\beta$ are integration constants. In the above theory space, it can be confirmed that the Hubble and scalar field equations become degenerate on the vacuum; both reduce to
\begin{equation}
\label{eq:HSeq_vac}
\left( \dfrac{2 A_X}{h} - \dfrac{A}{h X} - \dfrac{6 \sqrt{2} \alpha}{X} \right)\dot{X} +6 \left( A + 6 \sqrt{2} h \alpha \right) = 0 ,
\end{equation}
determining the kinetic density $X \left( t \right)$. However, the Friedmann constraint reduces to 
\begin{equation}
3 h ^2 = 6 \sqrt{2} h \alpha + \rho_\Lambda
\end{equation}
and shows that its \textit{explicit} time-dependence dropped out in going to the tadpole-free theory, i.e., no dynamical cancellation of vacuum energy and therefore \textit{no tempering}.

It is possible to proceed by disavowing the requirement of dynamical cancellation. However, it turns that the theory (Eq. \eqref{eq:almost_temp_theory}) will also suffer from either a ghost or gradient instability in vacuum.

\subsection{Self-tuning kinetic gravity braiding: a nontrivial tadpole-free limit}
\label{subsec:self_tuned_kgb}

The alternative to tempering is \textit{trivial scalar} self-tuning wherein the scalar field equation becomes trivially satisfied the de Sitter vacuum. The subset of KGB with this handy mechanism can be found by (1) taking the ``on-shell" de Sitter limit, $H\left( t \right) = h$, of the field equations (Eqs. \eqref{eq:friedmann_eq_tadpole_ds}, \eqref{eq:hubble_eq_tadpole_ds}, and \eqref{eq:sfe_tadpole_ds}) and (2) solving for the functional $B \left[ A \right]$ (Eq. \eqref{eq:b1}) that makes the scalar field equation identically satisfied. 

Step (1) has already been setup in the previous section (Eqs. \eqref{eq:friedmann_eq_tadpole_ds}, \eqref{eq:hubble_eq_tadpole_ds}, and \eqref{eq:sfe_tadpole_ds}). We continue by expressing the scalar field equation in terms of the kinetic density $X = \dot{\phi^2}/2$:
\begin{equation}
\dfrac{ \dot{X} A_X }{\sqrt{2 X}} + \sqrt{2 X} \left( 3 h A_X + \dot{X} A_{XX} \right) + 6 h \dot{X} B_X + 6 h X \left( 3 h B_X + \dot{X} B_{XX} \right) = l .
\end{equation}
It will be useful to recast this into the form
\begin{equation}
\left( A_X + 3 h \sqrt{2X} B_X \right) + \dfrac{ \dot{X} }{ 3 h \sqrt{2X} } \dfrac{d}{dX} \left[ \sqrt{2X} \left( A_X + 3 h \sqrt{2X} B_X \right) \right] = \dfrac{l}{3h\sqrt{2X}}
\end{equation}
and recognize the combination $A_X + 3 h \sqrt{2X} B_X$ as the dependent variable rather than $A\left(X\right)$ and $B\left(X\right)$ separately. Comparing the terms with and without the factor $\dot{X}$ on both sides of the equation leads to a differential equation and a constraint for $A_X + 3 h \sqrt{2X} B_X$:
\begin{equation}
\dfrac{d}{dX} \left[ \sqrt{2X} \left( A_X + 3 h \sqrt{2X} B_X \right) \right] = 0
\end{equation}
and
\begin{equation}
A_X + 3 h \sqrt{2X} B_X = \dfrac{l}{3h\sqrt{2X}} .
\end{equation}
[Step (2)] The differential equation is solved by $\sqrt{2X} \left( A_X + 3 h \sqrt{2X} B_X \right) = C$ where $C$ is an integration constant that can be fixed to $C = l/\left(3h\right)$ by the constraint equation. The exact solution to this system is therefore given by
\begin{equation}
\label{eq:b1}
A_X + 3 h \sqrt{2X} B_X = \dfrac{l}{ 3 h \sqrt{2X}} .
\end{equation}
Eq. \eqref{eq:b1} parametrizes the space of shift symmetric KGB with a self-tuning field of the trivial scalar type in the de Sitter state $H\left( t \right) = h$ \footnote{It is interesting that well-tempered KGB can be parametrized by
\begin{equation}
A_X + 3 h \sqrt{2X} B_X = \dfrac{l}{\sqrt{2X} \left( 3h + q\left( X \right) \right) }
\end{equation}
which coincides with Eq. \eqref{eq:b1} provided that $q \left(X\right)=0$.
}. It is straightforward to show that in the space of Eq. \eqref{eq:b1} the Friedmann constraint and Hubble equation reduce to
\begin{equation}
\label{eq:friedmann_eq_tadpole_st}
3h^2 = \rho - A - l \phi + \dfrac{ l \dot{\phi}}{3h} 
\end{equation}
and
\begin{equation}
\label{eq:hubble_eq_tadpole_st}
\rho + P + \dfrac{ l \dot{\phi} }{ 3 h } - \dfrac{l \ddot{\phi}}{9 h^2} + \dfrac{A_X \dot{\phi}\ddot{\phi}}{3h} = 0  ,
\end{equation}
respectively. Notably, by differentiating Eq. \eqref{eq:friedmann_eq_tadpole_st} with respect to the comoving time $t$, we obtain Eq. \eqref{eq:hubble_eq_tadpole_st} provided that the cosmic fluid satisfies
\begin{equation}
\label{eq:geodesics_tadpole_st}
\dot{\rho} + 3 h \left( \rho + P \right) = 0 .
\end{equation}
This is expected for any perfect fluid and so points to us an important distinction between tempering and trivial scalar self-tuning: tempering screens only vacuum energy and trivial scalar self-tuning blocks all forms of energy.

On the other hand, based on the Section \ref{subsec:well_tempered_kgb}'s no-tempering theorem, self-tuning within the tadpole-free theory can only be achieved by means of the trivial scalar type. In the next sections, we establish that self-tuning within the tadpole-free shift symmetric theory guarantees a dynamically stable and healthy late-time cosmic acceleration (Secs. \ref{subsec:stability_ds} and \ref{subsec:self_tuning_tailoring}).

Before moving on, we note that the implicit solution to Eqs. \eqref{eq:hubble_eq_tadpole_st} and \eqref{eq:geodesics_tadpole_st} for a barotropic perfect fluid with equation of state $P/\rho = w$, is given by
\begin{equation}
\label{eq:feq_st_vac}
3 h^2 = 3 h_0^2 e^{ - 3 h \left( 1 + w \right) t } - A \left( \dfrac{\dot{\phi}^2}{2} \right) - l \phi + \dfrac{ l \dot{\phi}}{3h}
\end{equation}
where $h_0$ is an integration constant. This is a first-order nonlinear differential equation which can be solved for $\phi \left( t \right)$ in the de Sitter vacuum. We note also that for $l = 0$ and $w = -1$ this reduces to an algebraic equation for $\dot{\phi} \left( t \right) = q$ where $q$ is a constant. In the nearly tadpole-free theory, $l \ll 1$, we can obtain a theory-agnostic Poincare asymptotic solution to Eq. \eqref{eq:feq_st_vac},
\begin{equation}
\label{eq:phi_asymp}
\phi \left( t \right) = \phi_0 + q t + \sum_{j = 1}^{N} \alpha_j \left( t \right) l^j .
\end{equation}
It is easy to obtain $\alpha_j \left( t \right)$ for any practical order in $l$. The first few $\alpha_j \left( t \right)$ are given by
\begin{equation}
6 h q A_X \alpha_1 \left( t \right) = -t \left(q (3 h t-2)+6 h \phi _0\right) ,
\end{equation}
\begin{equation}
\begin{split}
18 h^2 q^3 A_X^3 \alpha_2 \left( t \right) = & t A_X \left(q^2 (1-3 h t)-9 h^2 \phi _0^2\right) \\
& -q^2 t A_{XX} \left(q^2 (3 h t (h t-1)+1)+3 h \phi _0 \left(q (3 h t-2)+3 h \phi _0\right)\right) ,
\end{split}
\end{equation}
\begin{equation}
\begin{split}
648 h^3 q^5 A_X^5 \dfrac{ \alpha_3 \left( t \right) }{t} = & -3 q^4 A_{XX}^2 \left(q (3 h t-2)+6 h \phi _0\right) \\
& \times \left(q^2 (3 h t (3 h t-2)+2)+6 h \phi _0 \left(q (3 h t-2)+3 h \phi _0\right)\right) \\
& -36 h A_X^2 \left( 9 h^2 \phi _0^3+h q^3 t^2+q^2 \phi _0 (3 h t-1)\right) \\
& +q^2 A_X \bigg(q^2 A_{XXX} \left(3 h q t+6 h \phi _0-2 q\right) \\
& \times \left(q^2 (3 h t (3 h t-2)+2)+6 h \phi _0 \left(3 h q t+3 h \phi _0-2 q\right)\right) \\
& +3 A_{XX} \bigg(-108 h^3 \phi _0^3+q^3 (3 h t (h t (3 h t-16)+12)-8) \\
& +36 h q^2 \phi _0 (h t (h t-3)+1)\bigg) \bigg) ,
\end{split}
\end{equation}
where the potentials are evaluated at $X = q^2 / 2$, e.g., $A_X = A' \left( q^2 / 2 \right)$. In Sec. \ref{subsec:stability_ds}, we shall use this asymptotic solution to prove the dynamical stability of the de Sitter vacuum in the tadpole-free theory.

We elegantly express the gravitational action (Eq. \eqref{eq:kgb_st}) and field equations (Eqs. \eqref{eq:friedmann_eq_st}, \eqref{eq:hubble_eq_st}, and \eqref{eq:sfe_st}) of the self-tuning tadpole-driven shift symmetric KGB in terms of the single free potential $A\left(X\right)$,
\begin{equation}
\label{eq:kgb_st}
S_g = \int d^4 x \sqrt{-g} \left[ \dfrac{ R }{2} + l \phi + A \left( X \right) - \left( \dfrac{l}{18 h^2} \ln \left(X\right) - \dfrac{1}{3h} \int dX \dfrac{A_X}{ \sqrt{2X} } \right) \Box \phi \right] ,
\end{equation}
\begin{equation}
\label{eq:friedmann_eq_st}
3 H^2 = \rho - l \phi + \dfrac{l H \dot{\phi}}{3h^2} - A + \dot{ \phi }^2 \left( 1 - \dfrac{H}{h} \right) A_X ,
\end{equation}
\begin{equation}
\label{eq:hubble_eq_st}
2 \dot{H} = - \left( \rho + P \right) - \dfrac{l H \dot{\phi}}{3h^2} + \dfrac{l \ddot{\phi}}{9h^2} - \dot{\phi}^2 \left( 1 - \dfrac{H}{h} \right) A_X - \dfrac{\dot{\phi}\ddot{\phi}}{3h} A_X ,
\end{equation}
and
\begin{equation}
\label{eq:sfe_st}
\begin{split}
\left( 3H \left( 1 - \dfrac{H}{h} \right) - \dfrac{\dot{H}}{h} \right) \dot{\phi} A_X + \left( 1 - \dfrac{H}{h} \right) \left( A_X + \dot{\phi}^2 A_{XX} \right) \ddot{\phi} = l \left( 1 - \dfrac{H^2}{h^2} - \dfrac{\dot{H}}{3h^2} \right) .
\end{split}
\end{equation}
The above theory is valid for any $l$ and thus can be used to consistently obtain the tadpole-free limit ($l \rightarrow 0$). It is worth commenting that Eq. \eqref{eq:kgb_st} makes it clear that self-tuning KGB is described by only one free potential $A \left(X\right)$ instead of two, i.e., the price of requiring the scalar field to be self-tuning is the freedom of arbitrarily assigning two KGB potentials. Nonetheless, at this point, the free potential $A \left(X \right)$ remains to be completely arbitrary and can take on the form of any real function of a real variable. This will change when the gradient stability constraint is discussed. In the remainder of this section, we prove the stability of the de Sitter vacuum (Sec. \ref{subsec:stability_ds}), uncover an important dynamical implication of the tadpole-free theory (Sec. \ref{subsec:self_tuning_tailoring}), and derive ghost and gradient stability conditions (Sec. \ref{subsec:ghost_laplace_stability}). In Sec. \ref{sec:stgccg}, we provide numerical examples.

\subsection{Dynamical stability of the de Sitter vacuum}
\label{subsec:stability_ds}

To establish the dynamical stability of the de Sitter vacuum, $H \left( t \right) = h$, we define the perturbations $\left( f \left( t \right), \psi \left( t \right), \delta \rho \left( t \right) \right)$ to the Hubble parameter, scalar field, and the barotropic fluid density as
\begin{eqnarray}
H\left( t \right) &=& h + f \left( t \right) \\
\phi \left( t \right) &=& \bar{ \phi } \left( t \right) + \psi \left( t \right) \\
\rho \left( t \right) &=& \bar{ \rho} \left( t \right) + \delta \rho \left( t \right)
\end{eqnarray}
where $\bar{ \phi } \left( t \right)$ and $\bar{ \rho} \left( t \right)$ are the on-shell scalar field and fluid density, respectively. These perturbations describe tiny departures from the de Sitter vacuum and indicate whether the state $H \left( t \right) = h$ is dynamically stable or not (if perturbations diverge). Expanding the field equations (Eqs. \eqref{eq:friedmann_eq_st}, \eqref{eq:hubble_eq_st}, and \eqref{eq:sfe_st}) to first order in the perturbations, and making use of the background equations, we obtain, for $w \neq -1$ and $l \neq 0$,
\begin{equation}
\label{eq:dpsi}
\dot{\psi} -\dfrac{f \left(e^{3 h t (w+1)} \left(6 h^2 \ddot{ \bar{\phi} }-l \dot{ \bar{\phi} }^2\right)-9 \alpha ^2 h (w+1) \dot{ \bar{ \phi } }\right)}{h g \left( \bar{\phi}, t \right)}+\dfrac{\delta \rho e^{3 h t (w+1)} \ddot{ \phi } }{ g \left( \bar{\phi}, t \right) }-\dfrac{l \psi e^{3 h t (w+1)} \ddot{ \bar{ \phi } }}{g \left( \bar{\phi}, t \right)}=0 ,
\end{equation}
\begin{equation}
\label{eq:ddpsi}
\ddot{\psi} -\frac{6 h \dot{f} e^{3 h t (w+1)} \ddot{ \bar{ \phi } }}{ g \left( \bar{\phi}, t \right) }-3 f \dot{ \bar{ \phi } } - \dot{ \psi } \left(\dfrac{27 \alpha ^2 h^2 (w+1)^2}{ g \left( \bar{\phi}, t \right) }+\dfrac{ \dddot{ \bar{ \phi } } }{ \ddot{ \bar{ \phi } }}\right)-\dfrac{3 h (w+1) \delta \rho e^{3 h t (w+1)} \ddot{ \bar{\phi} } }{ g \left( \bar{\phi}, t \right) } = 0 ,
\end{equation}
and
\begin{equation}
\label{eq:df}
\dot{f}-\dfrac{f \left(9 \alpha ^2 h (w+1) \left(3 h w \ddot{ \bar{ \phi } } + \dddot{ \bar{ \phi } } \right)-l e^{3 h t (w+1)} \left( \dot{ \bar{ \phi } } \left(3 h \ddot{ \bar{ \phi } }- \dddot{ \bar{ \phi } }\right)+2 \ddot{ \phi }^2\right)\right)}{ \ddot{ \bar{ \phi } } g \left( \bar{\phi}, t \right)  }=0 ,
\end{equation}
where $\alpha$ is defined by the background fluid density, $\bar{ \rho} = 3 \alpha^2 e^{-3 h t  \left( 1 + w \right)}$, and
\begin{equation}
\label{eq:g_def}
g \left ( \bar{\phi}, t \right) = 9 h \left( 1 + w \right) \alpha ^2 + l \dot{ \bar{\phi}} e^{3 h t \left( 1 + w \right) } .
\end{equation}
Eq. \eqref{eq:dpsi} is a constraint equation to Eqs. \eqref{eq:ddpsi} and \eqref{eq:df}. In addition to this, the (barotropic) density perturbation $\delta \rho$ satisfies the same equation as the background fluid \footnote{For nonbarotropic fluid perturbations, e.g., $\delta P = w \delta \rho - \left( S / 3 h \right)$, we could just include a source term $S \left( t \right)$ in the right hand side of Eq. \eqref{eq:drho}.}
\begin{equation}
\label{eq:drho}
\dot{ \delta \rho } + 3 h \left( 1 + w \right) \delta \rho = 0 
\end{equation}
and so we write down
\begin{equation}
\delta \rho = 3 \delta h^2 e^{ -3 h t \left( 1 + w \right) }
\end{equation}
where $\delta h$ is an integration constant. Provided the background scalar field $\bar{ \phi }$ and the fluid equation of state $w$, the two other perturbations, $\psi$ and $f$, can be obtained by first solving Eq. \eqref{eq:df} for the Hubble parameter's perturbation $f$, and then solving Eq. \eqref{eq:ddpsi} for the scalar field's perturbation $\psi$. The integration constants can be fixed using the constraint equation (Eq. \eqref{eq:dpsi}). However, in general, the explicit expression for the background scalar field $\bar{ \phi }$ can be obtained only after specifying the potential $A\left(X\right)$ in Eq. \eqref{eq:friedmann_eq_tadpole_st}.

Some physical considerations can be made to make progress at this stage even without specifying an exact model. The cosmic fluid is composed of noninteracting species, each described by a density $\rho_i$ and pressure $P_i$, which move on the geodesics of the background. However, the cosmic expansion draws energy from of each of these species, except for vacuum energy $\left( w = -1 \right)$ which eventually would leave the expansion accelerating. Thus, vacuum energy ($w = -1$), regardless of its magnitude, would be the only source of the expansion at late times \footnote{This of course does not count the scalar field $\phi$ which cannot be considered as a perfect fluid because of the braiding term \cite{st_galileon_inflation_deffayet}.}. Furthermore, in the tadpole-free limit ($l = 0$), the background and perturbations can be solved exactly independent of the model. 

To obtain the perturbed field equations for $w = -1$ and $l = 0$, we set $w = -1$ in Eqs. \eqref{eq:friedmann_eq_st}, \eqref{eq:hubble_eq_st}, and \eqref{eq:sfe_st}, expand about the perturbations $\left( f \left( t \right), \psi \left( t \right), \delta \rho \left( t \right) \right)$, and use the Poincare asymptotic expansion given by Eq. \eqref{eq:phi_asymp}. The resulting expressions are given by
\begin{equation}
\dot{\psi} -\dfrac{h \delta \rho_\Lambda - f \left(q^2 A_X + 6 h^2\right)}{h q A_X}=0 ,
\end{equation}
\begin{equation}
\ddot{\psi} + \dfrac{6 h \dot{f}}{q A_X}-3 q f=0 ,
\end{equation}
and
\begin{equation}
\dot{f} + 3 h f = 0 ,
\end{equation}
where $\delta \rho_\Lambda$ is the perturbation to vacuum energy density and the potentials are evaluated at $X = q^2 / 2$. This coupled system can easily be solved, leading to the perturbations
\begin{eqnarray}
\label{eq:h_asymp} f &=& \delta H e^{-3 h t} \\
\label{eq:psi_asymp} \dot{\psi} &=& - \dfrac{ \delta H }{h} \left( q + \dfrac{6 h^2}{q A_X} \right) e^{-3 h t} + \dfrac{\delta \rho_\Lambda}{q A_X} 
\end{eqnarray}
where $\delta H$ is an integration constant. Because of shift symmetry, we present $\dot{\psi}$ instead of $\psi$ itself. Fixing the constant $\delta \rho_\Lambda$ to zero by recalibrating the vacuum energy density $\rho_\Lambda$, we see that the perturbations vanish at $t \rightarrow \infty$. This proves the dynamical stability of the de Sitter state in the tadpole-free self-tuning theory. The numerical integrations in Sec. \ref{sec:stgccg} will support this conclusion.

\subsection{Self-tuning and the vanishing current-hypersurface}
\label{subsec:self_tuning_tailoring}

In this section, we reveal an important connection between self-tuning and the dynamical hypersurface defined by
\begin{equation}
J \left( t \right) = 0
\end{equation}
where
\begin{equation}
\label{eq:j_kgb}
J \left( t \right) = \dot{\phi} K_X + 3 H \dot{\phi}^2 G_X - 2 \dot{\phi} G_\phi .
\end{equation}
The quantity $J \left( t \right)$ could be recognized as a conserved Noether current arising from shift symmetry. First, we show that in the tadpole-free shift symmetric theory the late-time dynamics of cosmologically interesting expansion histories reside on the hypersurface $J \left( t \right) = 0$. We then show that in the self-tuning theory this hypersurface coincides with the de Sitter state, $H\left(t\right) = h$, suggesting that cosmological dynamics would exit at late-time cosmic acceleration.

To get to the first point, we note that the scalar field equation for the shift symmetric theory can be written down as
\begin{equation}
\dfrac{1}{a \left( t \right)^3} \dfrac{d}{dt} \left( a \left( t \right)^3 J \left( t \right) \right) = l .
\end{equation}
The solution to this is given by
\begin{equation}
J \left( t \right) = \dfrac{C}{a\left( t \right)^3} + \dfrac{l}{a\left(t\right)^3} \int^t d y \ a \left( y \right) ^3
\end{equation}
where $C$ is an arbitrary constant. For the power law $\left( a = \left( t/T \right)^n\right)$ and exponential $\left( a = a_0 e^{-3 H_0 t} \right)$ expansion histories the scalar current reduces to
\begin{equation}
J \left( t \right) = C \left( \dfrac{T}{t} \right)^{3n} + \dfrac{ l t }{1 + 3n}
\end{equation}
and
\begin{equation}
J \left( t \right) = \dfrac{C}{a_0^3} e^{-3 H_0 t} + \dfrac{l}{3H_0} ,
\end{equation}
respectively. This shows that in expanding universes ($n > 0$ and $H_0 > 0$) the late-time form of the scalar current can be written as $J \left( t \right) \sim l \left( a + b t \right)$ where $a$ and $b$ are finite constants. For the tadpole-free theory, the cosmological dynamics would instead end up on $J\left( t \right) = 0$ for any expanding universes. This is interesting considering that nearly arbitrary cosmological dynamics can be accommodated in this hypersurface \cite{bernardo2019tailoring, designing_horndeski_arjona}.

We now show that in the self-tuning theory (Sec. \ref{subsec:self_tuned_kgb}) the hypersurface $J\left( t \right) = 0$ coincides with the de Sitter vacuum, $H\left( t \right) = h$. This becomes transparent by writing down the conserved current (Eq. \eqref{eq:j_kgb}) in the self-tuning theory:
\begin{equation}
\label{eq:conserved_current_j_st}
J \left( t \right) = \left( 1 - \dfrac{H}{h} \right) \dot{\phi} A_X + \dfrac{l H}{3 h^2} .
\end{equation}
For $l = 0$, we see that the de Sitter vacuum, $H\left( t \right) = h$, coincides with the vanishing current hypersurface $J \left( t \right) = 0$. Combining this with the earlier observation that cosmological dynamics would always end up on $J \left( t \right) = 0$ in expanding universes, we reach an interesting conclusion: cosmological dynamics in the tadpole-free self-tuning theory would always exit at cosmic acceleration. This is a remarkable output of self-tuning in the tadpole-free shift symmetric KGB.

We warn however that the de Sitter vacuum is not the only unique state on the hypersurface $J\left( t \right) = 0$. By looking again at the $l = 0$ limit of Eq. \eqref{eq:conserved_current_j_st}, we see that $J\left( t \right) = 0$ can also be obtained provided that $\dot{\phi} A_X = 0$. This alternative late-time state corresponds to the vanishing of the scalar field's stress-energy tensor. The perturbations reveal that the state $\dot{\phi} A_X = 0$ suffers from ghost and gradient instabilities; moreover, the dark energy equation of state strongly distinguishes this late-time limit from the de Sitter vacuum.

\subsection{Ghost and Laplace stability conditions}
\label{subsec:ghost_laplace_stability}

In this section, we show that the self-tuning theory can be strongly constrained by imposing ghost and gradient stability conditions on the vacuum. In KGB, it can be shown that ghost and gradient instabilities can be avoided provided that
\begin{eqnarray}
\mathcal{F} &>& 0 \\
\mathcal{G} &>& 0 
\end{eqnarray}
where
\begin{equation}
\mathcal{F} = K_X + 2 \left( \ddot{\phi} + 2H \dot{\phi} \right) G_X - \dfrac{\dot{\phi}^4}{2} G_X^2 + \dot{\phi}^2 \ddot{\phi} G_{XX} - 2  G_\phi + \dot{\phi}^2 G_{\phi X}
\end{equation}
and
\begin{equation}
\mathcal{G} = K_X + \dot{\phi}^2 K_{XX} + 6 H \dot{\phi} G_X + \dfrac{3}{2} \dot{\phi}^4 G_X^2 - 2 G_\phi - \dot{\phi}^2 G_{\phi X} + 3 H \dot{\phi}^3 G_{XX} .
\end{equation}
These conditions can be obtained by evaluating the second-order action for perturbations in the unitary gauge $\left( \delta \phi = 0 \right)$ and making sure that the signs of the kinetic and gradient terms are correct \cite{st_galileon_inflation_deffayet, st_galileon_inflation_kobayashi, st_horndeski_cubic_appleby}. For the self-tuning theory, the functions $\mathcal{F}$ and $\mathcal{G}$ reduces to
\begin{equation}
\mathcal{F} = -\dfrac{\sqrt{2} \sqrt{X} A_{XX} \ddot{ \phi }}{3 h}+A_X \left(\frac{\sqrt{2} l \sqrt{X}}{27 h^3}-\frac{4 H \dot{ \phi } + \ddot{ \phi }}{3 \sqrt{2} h \sqrt{X}}+1\right)-\frac{X A_X^2}{9 h^2}+\frac{l \left(36 h^2 H \dot{ \phi } - l X\right)}{162 h^4 X}
\end{equation}
and 
\begin{equation}
\mathcal{G} = A_{XX} \left(2 X-\frac{\sqrt{2} \sqrt{X} H \dot{ \phi } }{h}\right)+A_X \left(-\frac{\sqrt{2} l \sqrt{X}}{9 h^3}-\frac{H \dot{\phi} }{\sqrt{2} h \sqrt{X}}+1\right)+\frac{X A_X^2}{3 h^2}+\frac{l^2}{54 h^4} ,
\end{equation}
respectively. These conditions should be satisfied by a cosmological background solution to be free from ghost $\left( \mathcal{G} > 0 \right)$ and gradient $\left( \mathcal{F} > 0 \right)$ instabilities. The late-time state $\dot{\phi} A_X = 0$ in the $l = 0$ subclass violates these stability constraints.

Turning our attention to the de Sitter vacuum in tadpole-free theory, characterized by $H = h$ and $\dot{\phi} = q$, the stability conditions reduce to 
\begin{equation}
-\dfrac{A_X \left(q^2 A_X+6 h^2\right)}{18 h^2}>0
\end{equation}
and
\begin{equation}
\frac{q^2 A_X^2}{6 h^2}>0
\end{equation}
where the $A_X$ is evaluated at $X = q^2/2$. The latter condition is obviously already trivially satisfied. On the other hand, the former (gradient stability constraint) can be satisfied when $A_X < 0$ and $ q^2 A_X+6 h^2 > 0$. In summary, the ghost and gradient stability of the de Sitter vacuum under perturbations can be ensured for theory potentials $A\left( X \right)$ which satisfy
\begin{equation}
\label{eq:ghost_grad_stab_st}
- \dfrac{3 h^2}{X} < A_X < 0 .
\end{equation}
This shows that viable tadpole-free shift symmetric KGB lives only in the very tight space described by Eq. \eqref{eq:ghost_grad_stab_st}.

We note that the unhealthy de Sitter state given covariantly by $A_X = 0$ or at component level by the equation $3 H^2 = \rho_\Lambda$ can be identified as a stealth solution wherein the background field equations reduce merely to that of GR \cite{stealth_bernardo}. The ghost can then be traced to the same one surrounding stealth black holes and can be alleviated using the \textit{scordatura} mechanism \cite{st_bh_pt_de_rham, scordatura_motohashi, scordatura_gorji}. This is outside the scope of this paper but is very interesting to deserve being mentioned.

\section{Self-tuning generalized cubic Galileon}
\label{sec:stgccg}

In this section, we apply the results of the previous section to the generalized cubic covariant Galileon. The resulting cosmological theory possesses two de Sitter attractors, one of which is the self-tuning vacuum (Sec. \ref{subsec:gccg}) which we further explore in several numerical experiments involving a mixture of cosmic fluids (Secs. \ref{subsec:dynamical_system_analysis} and \ref{subsec:numerical_experiments}). In addition, we also explore the stability of the vacuum under phase transitions (Sec. \ref{subsec:phase_transition}). We conclude by addressing the viability of tuning away an arbitrarily large bare cosmological constant (Sec. \ref{subsec:self_tuning_gccg_vs_large_cc}).

\subsection{Generalized cubic covariant Galileon}
\label{subsec:gccg}

The generalized cubic covariant Galileon theory is given by the $k$-essence and braiding potentials
\begin{eqnarray}
K \left( \phi, X \right) &=& c X^n \\
G \left( \phi, X \right) &=& d X^m
\end{eqnarray}
where $c$, $d$, $n$, and $m$ are constants. The Galileon limit ($n = 1$, $m = 1$) of this theory has been very much studied owing to properties such as Vainshtein screening and the existence of a tracker solution, $H \dot{\phi} = \text{constant}$ \cite{st_horndeski_galileon_barreira_2, st_horndeski_renk, st_horndeski_galileon_peirone}. The tracker of cubic Galileon, however, predicts a dark energy equation of state away from $w_{\phi} = -1$ at the matter era \cite{st_dark_energy_tsujikawa}. For the generalized cubic covariant Galileon, a phenomenologically viable tracker solution of the form $H \dot{\phi}^{2q} = \text{constant}$ persists when $m = n + (2 q - 1)/2$ \cite{st_horndeski_cosmology_de_felice}.

Self-tuning in the generalized cubic covariant Galileon constrains the exponent of the braiding potential to $m = n - 1/2$. The case $n = 1$ of this has been studied in Ref. \cite{self_tuning_bh_emond} in the context of cosmological black holes. It should also be mentioned that the resulting theory loses the tracker solution ($q = 0$). The potentials of the self-tuning theory are given by
\begin{eqnarray}
\label{eq:k_gccg_st}
K \left( \phi, X \right) &=& c X^n \\
\label{eq:g_gccg_st}
G \left( \phi, X \right) &=& - \dfrac{c n X^{n - 1/2}}{3 \sqrt{2} h \left( n - 1/2 \right)} .
\end{eqnarray}
Furthermore, the gradient stability constraint on the de Sitter vacuum, $H \left( t \right) = h$, of the theory leads to
\begin{equation}
\label{eq:ngc}
- \dfrac{1}{\left( \rho_\Lambda /\left( 3h^2 \right) - 1 \right)} < n < 0 .
\end{equation}
This shows that a healthy self-tuning generalized cubic covariant Galileon theory must have singular potential and so may not even admit a Minkowski vacuum. The field equations (Eqs. \eqref{eq:friedmann_eq_st}, \eqref{eq:hubble_eq_st}, and \eqref{eq:sfe_st}) become
\begin{equation}
\label{eq:friedmann_eq_gccg}
c \left(-2 h n+h+2 n H \right) \dot{\phi}^{2n}+h 2^n \left(3 H^2-\rho\right) = 0 ,
\end{equation}
\begin{equation}
\label{eq:hubble_eq_gccg}
2 c n \dot{\phi}^{2n - 1} \left(3 (h-H) \dot{ \phi } + \ddot{ \phi } \right) +3 h 2^n \left(2 \dot{H}+P+\rho \right)=0 ,
\end{equation}
and 
\begin{equation}
\label{eq:sfe_gccg}
\dot{\phi}^{2n-1} \left( \dot{\phi} \left(3 H ( h - H ) - \dot{H} \right) + ( 2 n - 1 ) ( h - H ) \ddot{ \phi } \right) = 0.
\end{equation}
Clearly, the scalar field equation (Eq. \eqref{eq:sfe_gccg}) is satisfied on the de Sitter vacuum, $H \left( t \right) = h$. In this state $\ddot{\phi} = 0$ also follows by virtue of Eq. \eqref{eq:hubble_eq_gccg} and so the cancellation of the bare vacuum energy can be viewed as the entry of arbitrary integration constants in the Hamiltonian constraint (Eq. \eqref{eq:friedmann_eq_gccg}), $3 h^2 = \rho_\Lambda -  c \left( \dot{\phi}_0^2 / 2 \right)^n$. This scenario holds generally with $\dot{\phi} \left( t \right)$ being a constant (i.e., $\phi \sim t$ on-shell) in tadpole-free models and otherwise explicitly time-dependent in tadpole-driven models. In both cases, the time dependence brought by $\phi \left( t \right)$ effectively reduces to arbitrary integration constants in the Hamiltonian constraint, counteracting the bare vacuum energy $\rho_\Lambda$. On the other hand, as mentioned previously, there is another de Sitter attractor given by $3 H^2 = \rho_\Lambda$ where $\rho_\Lambda$ is the bare vacuum energy. This attractor is unhealthy and comes with a diverging scalar field, $\dot{\phi} \sim \exp \left( \sqrt{3 \rho_\Lambda} t / \left( 1 - 2n \right) \right)$, for $n < 0$. It can also be shown that it comes with a dark energy equation of state $w_\phi = -1 / \left( 1 - 2n \right)$. This will be confirmed using numerical experiments.

It may be worth pointing out that there is a self-tuning theory with $n = 1/2$. The apparent break down of Eq. \eqref{eq:g_gccg_st} for $n = 1/2$ is an artifact of the power-law parametrization of the $k$-essence potential (Eq. \eqref{eq:k_gccg_st}). By using $K \left( \phi, X \right) = c \sqrt{X}$ and using Eq. \eqref{eq:b1}, one easily ends up with $G\left(\phi, X\right) = -c \ln\left(X\right)/\left( 6\sqrt{2} h \right)$. This theory can be shown to be incompatible with the gradient stability constraint; nonetheless, it completes the analysis above and serves as an independent example.

\subsection{Dynamical system analysis with single non-$\Lambda$ fluid}
\label{subsec:dynamical_system_analysis}

We start our numerical experiments first with only a single non-$\Lambda$ cosmic fluid. In this case, we perform a dynamical system analysis to reveal that the attractors of the theory live in the hypersurface $J\left( t \right) = 0$ and also confirm this by numerically integrating the field equations.

The scalar current of the theory is given by
\begin{equation}
\label{eq:current_gccg}
J = 2^{1 - n} c n \left( 1 - \dfrac{H}{h} \right) \dot{\phi}^{2n - 1} .
\end{equation}
We introduce the cosmic fluids via $\rho \rightarrow \rho + \rho_\Lambda$ and $P \rightarrow w \rho - \rho_\Lambda$ where $w$ is the equation of state of the barotropic non-$\Lambda$ fluid, e.g., $w = 0$ for baryons and dark matter, $w = 1/3$ for photons and neutrinos. To obtain the dynamical system, we eliminate $\rho$ in the Hubble equation (Eq. \eqref{eq:hubble_eq_gccg}) by using the Friedmann constraint (Eq. \eqref{eq:friedmann_eq_gccg}). In this way, we end up with the two-dimensional dynamical system for the vector $\left( J , H \right)$:
\begin{equation}
\dot{J} = - 3 H J 
\end{equation}
and
\begin{equation}
\begin{split}
\dot{H}= - \dfrac{3 \left( h-H \right)}{D\left( t \right)} \bigg[
& -h 2^n (2 n-1) (w+1) \left(\rho_\Lambda -3 H^2\right) \\
&-c ((2 n-1) w-1) (h (2 n-1)-2 n H) \\
&\times \left(cn\right)^{2n/(1-2n)} 2^{2n(1-n)/(1-2n)} \left(\dfrac{h-H}{h J}\right)^{2n/(1-2 n)} \bigg] .
\end{split}
\end{equation}
where
\begin{equation}
D \left( t \right) = 2 \left( \left( c n \right)^{1/\left(1-2n\right)} 2^{2n(1-n)/(1-2n)} \left( \dfrac{h-H}{h J} \right)^{2n/\left(1-2 n\right)}+3 h 2^n (2 n-1) (h-H)\right) .
\end{equation}

A few comments about this dynamical system are in order. First, for healthy theories, $n < 0$, it is easy to check that there are two fixed points along the $H$ axis ($J = 0$). These are given by $H = h$ (on-shell) and $3 H^2 = \rho_\Lambda$. However, it must be mentioned that to approach the on-shell vacuum $J$ must vanish faster than $H$ approaches $h$. Second, in general, the vector field $\left( \dot{J}, \dot{H} \right)$ will be real only for $H > h$. The exception is for very special cases where $2n/(1 - 2n)$ reduces to a positive integer. But these cases are outside the scope of healthy theories ($n < 0$). Third, along the curve $H_C \left[ J \right]$ defined by $D \left( t \right) = 0$ the vector field $\left( \dot{J}, \dot{H} \right)$ diverges. As will be demonstrated in a short while, this curve bifurcates the phase space of the dynamical system depending on what de Sitter state will be reached at late times. We also emphasize that the blow up of the vector field $\left( \dot{J}, \dot{H} \right)$ on the curve $H_C \left[ J \right]$ is a coordinate artifact. To study the dynamics along the curve, we go to the coordinate system $\left( \dot{\phi}, H \right)$. In this case, the vector field $\left( \ddot{\phi}, \dot{H} \right)$ continues to blow up on $H_C \left[ \dot{\phi} \right] = H_C \left[ J \left[ \dot{\phi}, H \right] \right]$ (see Eq. \eqref{eq:current_gccg}). Nonetheless, the quantity $dH/d\dot{\phi}$ remains finite everywhere including points on $H_C \left[ \dot{\phi} \right]$. Evaluating $dH/d\dot{\phi}$ on $H_C \left[ \dot{\phi} \right]$ we obtain the reduced (one-dimensional) dynamical system given by
\begin{equation}
\label{eq:ds_H_C}
dH/d\dot{\phi} \bigg |_{H_C} = - \dfrac{c n}{3 \times 2^n h} \dot{\phi}^{2n - 1} .
\end{equation}
This equation captures the dynamics on $H_C \left[ J \right]$. For healthy theories ($n < 0$), however, the overall coefficient of the (one-dimensional) vector $dH/d\dot{\phi} |_{H_C}$ is positive and so the system has only an unstable fixed point. This fixed point is located either at $H = h$ (with $\dot{\phi} = 0$) if $n > -1/2$ or at $3H^2 = \rho_\Lambda$ (with $\dot{\phi} = \infty$) if $n < -1/2$. Lastly, it is possible to further use the arclength on the $\left( \dot{\phi}, H\right)$-coordinate system as an independent variable to obtain a dynamical system that is finite everywhere. In this nonsingular coordinate system the de Sitter states are no longer fixed points. For this paper, we prefered to use the variables $J$ (or $\dot{\phi}$) and the comoving time $t$ because we wanted to show that the cosmological dynamics always relaxes to $J\left( t \right) = 0$.

To nondimensionalize the scalar field we set $\left[ c \right] = \left[ T \right]^{2 (n - 1)}$ where $\left[ T \right]$ is a unit of time. By using the transformations
\begin{eqnarray}
\beta &=& c h^{2(n - 1)} \\
\tau &=& h t \\
x \left( \tau \right) &=& J \left( \tau/ h \right) / h \\
y \left( \tau \right) &=& H \left( \tau/ h \right) / h \\
\lambda &=& \rho_\Lambda / 3h^2 
\end{eqnarray}
we obtain the dimensionless dynamical system given by
\begin{equation}
\label{eq:dx}
x' = -3 x y
\end{equation}
and
\begin{equation}
\label{eq:dy}
\begin{split}
y' =
- \dfrac{ 3 (1-y) }{d\left( \tau \right)} \bigg[ 
& 3 \times 2^n (2 n-1) (w+1) \left(y^2-\lambda \right) \\
& + \beta  ((2 n-1) w-1) (2 n y-2 n+1) \\
& \times \left( \beta n \right)^{2n/(1-2n)} 2^{2n(1-n)/(1-2n)} \left( \dfrac{ 1 - y }{x}\right)^{2n/(1-2n)} \bigg]
\end{split}
\end{equation}
where a prime denotes differentiatiation with respect to $\tau$ and
\begin{equation}
d \left( \tau \right) = 2 \left( \left( \beta n \right)^{1/(1-2n)} 2^{2n(1-n)/(1-2n)} \left(\dfrac{ 1 - y }{ x }\right)^{ 2n/(1-2n) }-3 \times 2^n (2 n-1) (y-1)\right) .
\end{equation}
In this system, the on-shell de Sitter state is given by $y = 1$ and the other de Sitter limit is $y^2 = \lambda$.

\begin{figure}[h!]
\center
	\subfigure[ ]{
		\includegraphics[width = 0.48 \textwidth]{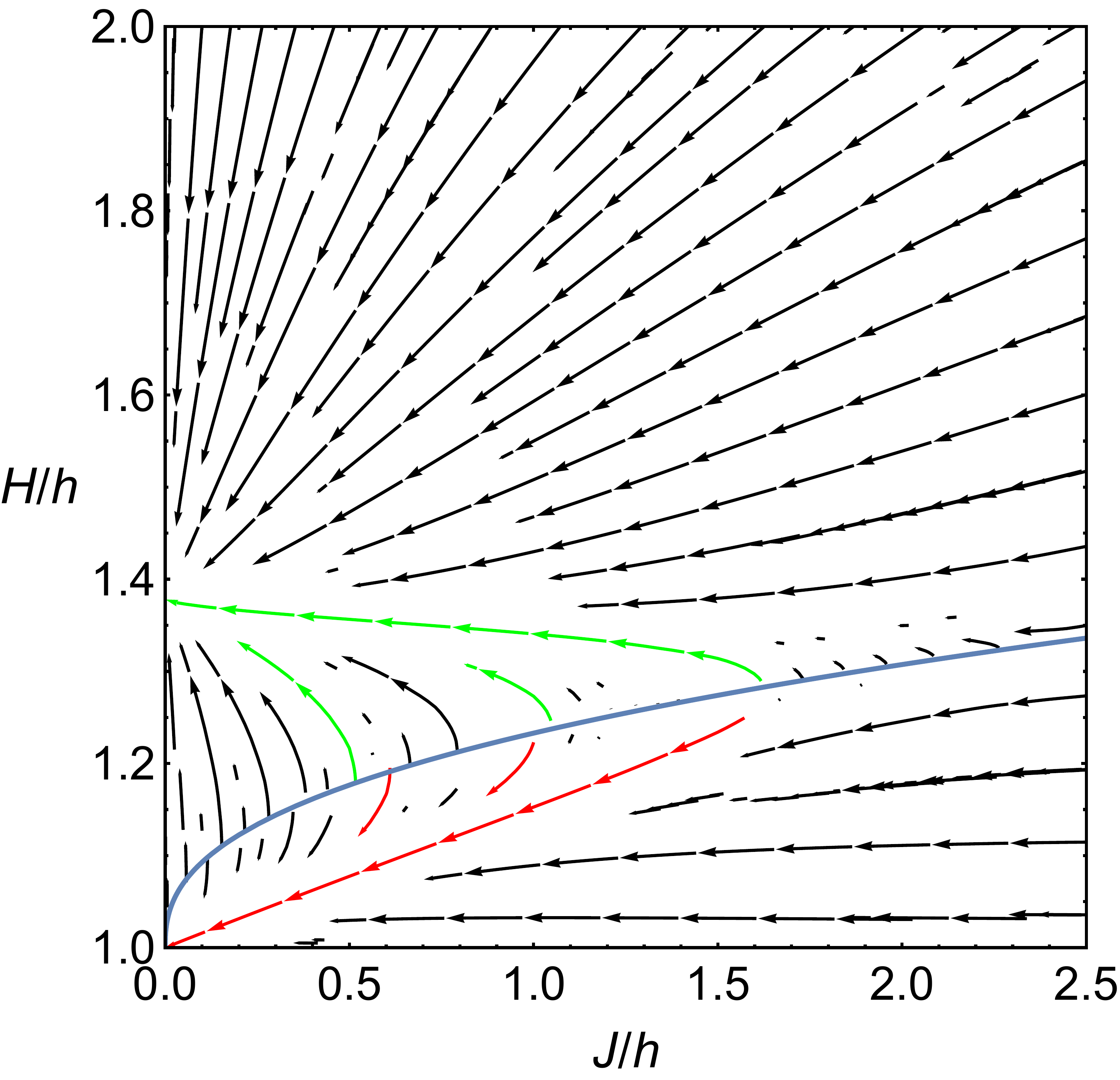}
		}
	\subfigure[ ]{
		\includegraphics[width = 0.44 \textwidth]{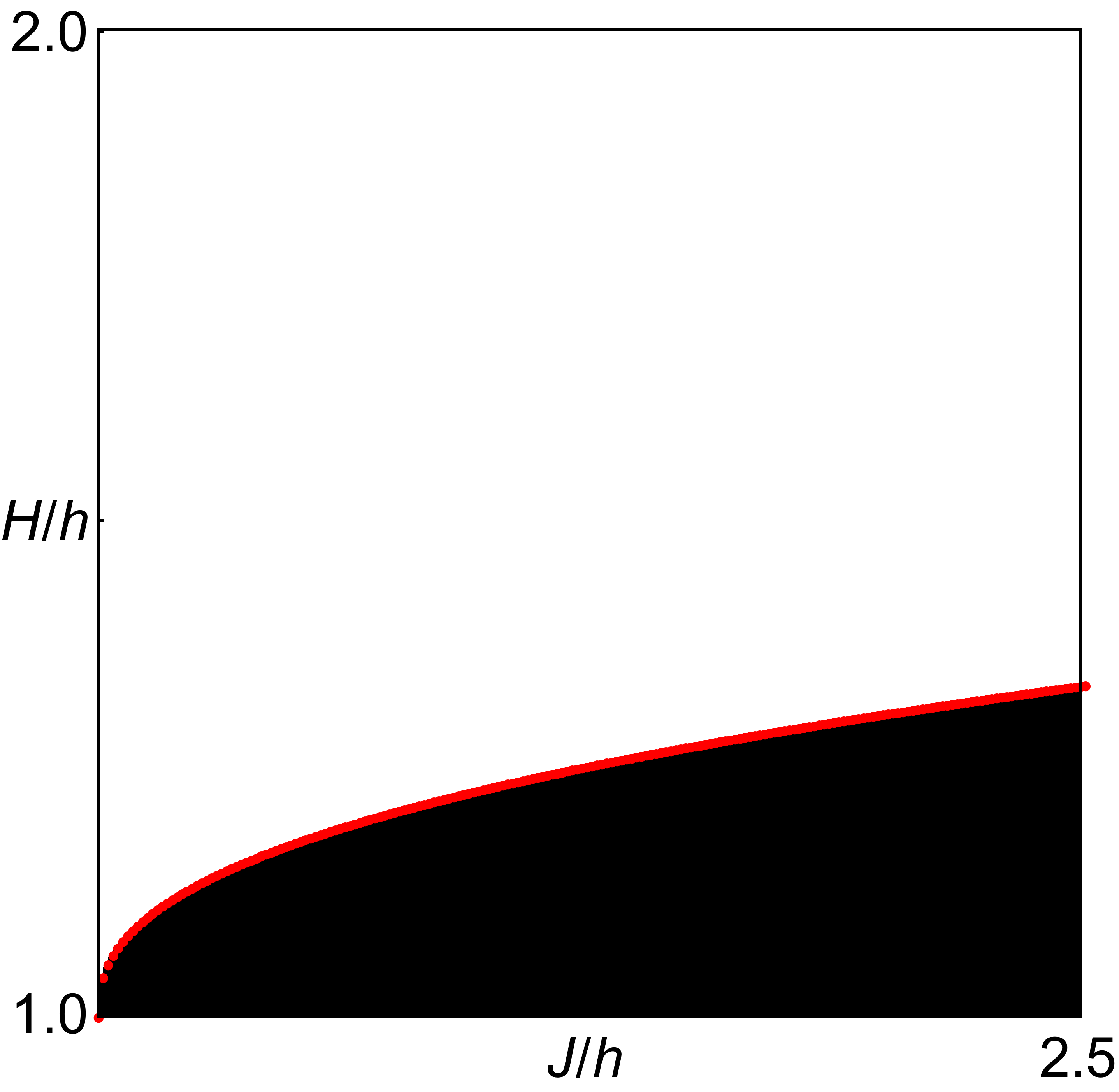}
		}
\caption{$\left[ n = -1 , \lambda = 19/10, w = 0 , \beta = 1 \right]$ (a) Phase portrait of the vector field $(x', y')$ given by Eqs. \eqref{eq:dx} and \eqref{eq:dy}. Blue solid line is $H_C\left[J\right]$ and red (green) stream lines come from points slightly below (above) $H_C\left[J\right]$. (b) $200 \times 200$ pixels-plot of phase space obtained by numerically integrating Eqs. \eqref{eq:dx} and \eqref{eq:dy} and characterizing the late-time states. Red markers show $H_C\left[J\right]$ and black (white) region shows the section of phase space which falls to the healthy $y = 1$ (unhealthy $y^2 = \lambda$) asymptotic state.}
\label{fig:ds1}
\end{figure}

\begin{figure}[h!]
\center
	\subfigure[ ]{
		\includegraphics[width = 0.48 \textwidth]{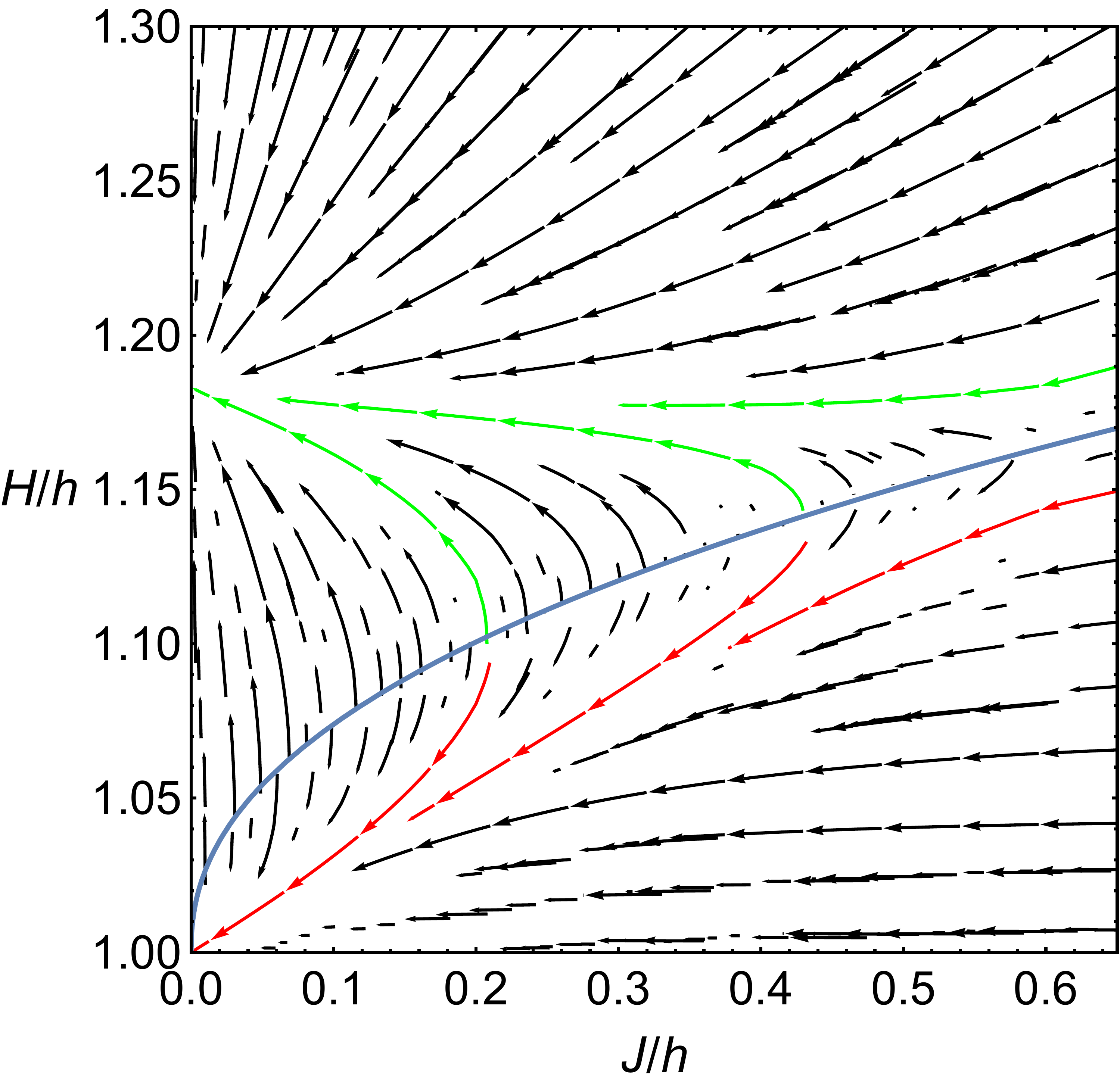}
		}
	\subfigure[ ]{
		\includegraphics[width = 0.46 \textwidth]{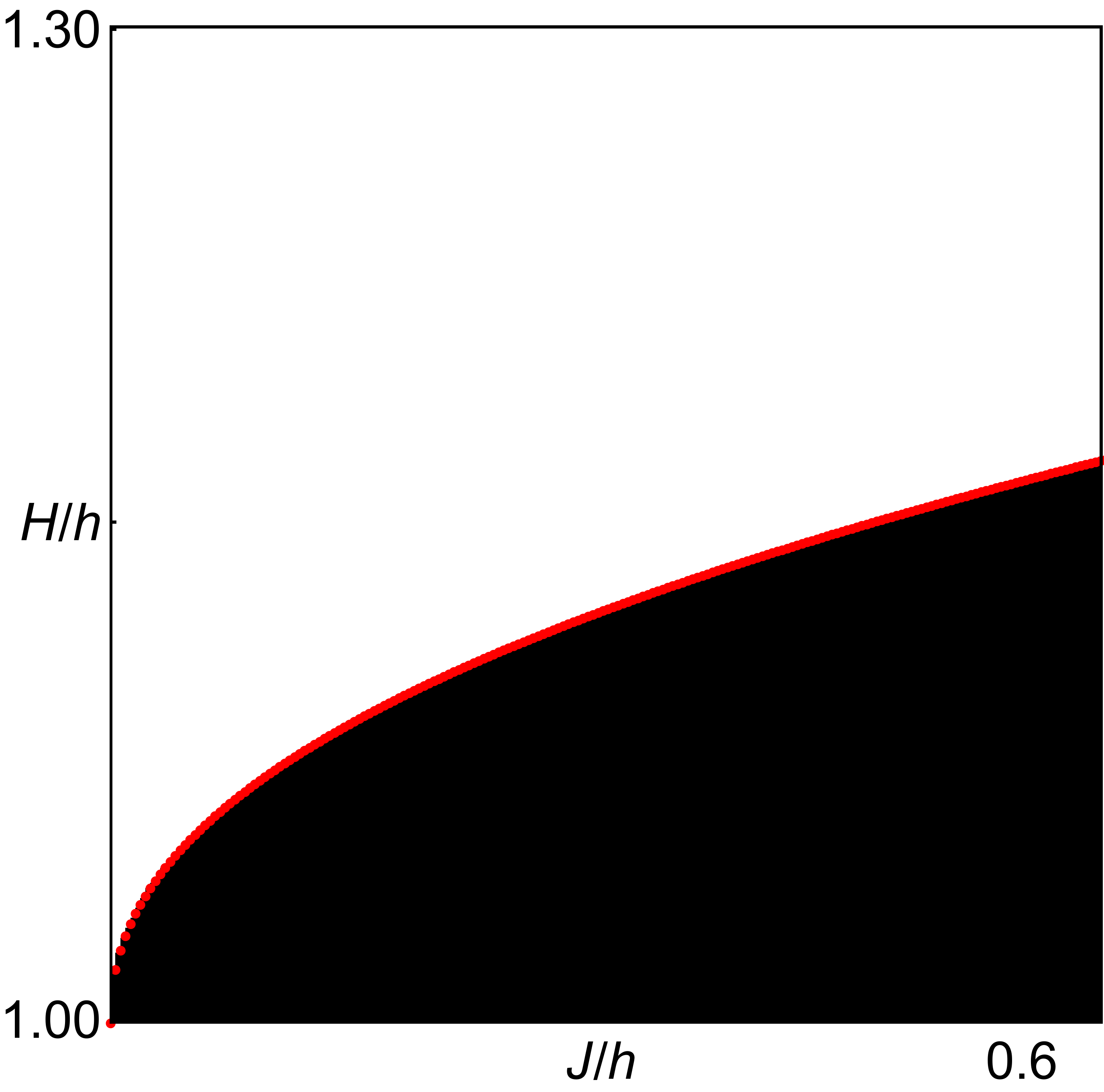}
		}
\caption{$\left[ n = -2 , \lambda = 7/5, w = 0 , \beta = 1 \right]$ (a) Phase portrait of the vector field $(x', y')$ given by Eqs. \eqref{eq:dx} and \eqref{eq:dy}. Blue solid line is $H_C\left[J\right]$ and red (green) stream lines come from points slightly below (above) $H_C\left[J\right]$. (b) $200 \times 200$ pixels-plot of phase space obtained by numerically integrating Eqs. \eqref{eq:dx} and \eqref{eq:dy} and characterizing the late-time states. Red markers show $H_C\left[J\right]$ and black (white) region shows the section of phase space which falls to the healthy $y = 1$ (unhealthy $y^2 = \lambda$) asymptotic state.}
\label{fig:ds2}
\end{figure}

We present the numerical experiments with a single non-$\Lambda$ cosmic fluid. For $\{ n = -1 , \lambda = 19/10, w = 0, \beta = 1 \}$, figure \ref{fig:ds1}(a) shows the phase portrait of the vector field $\left( x',y' \right)$. We stress out again that $y = H/h >1$; otherwise, the vector field $\left( x',y' \right)$ will have an imaginary component. This phase portrait confirms that the dynamical system has two stable fixed points which are $y = 1$ and $y^2 = \lambda$.
Also, shown in figure \ref{fig:ds1}(a) is the curve $H_C \left[ J \right]$ (blue solid line) which partitions the phase space depending on the late-time dynamics. This is confirmed for stream lines coming from points slightly above $H_C \left[ J \right]$ (green arrows) which fall straight to $y^2 = \lambda$ and for points slightly below $H_C \left[ J \right]$ (red arrows) which go to $y = 1$. This conclusion is further supported by numerically integrating the dynamical system and checking out the late-time state. This brings us to figure \ref{fig:ds1}(b) which is a $200 \times 200$ pixels plot of phase space which separates the regions ending up on $y = 1$ (black) from regions ending up on $y^2 = \lambda$ (white). The red line in figure \ref{fig:ds1}(b) is again the curve $H_C \left[ J \right]$ which partitions the phase space. The viable phase space of this theory is of course those which fall to the healthy on-shell limit $y = 1$. Very similar plots can be obtained for the radiation-dominated case $w = 1/3$. We encourage the reader to test this statement using the supplementary codes. For $\{ n = -2 , \lambda = 7/5, w = 0, \beta = 1 \}$, similar conclusions have been obtained as shown in figures \ref{fig:ds2}(a) and \ref{fig:ds2}(b).
Again, the phase portrait (figure \ref{fig:ds2}(a)) reveals the two de Sitter attractors and this is supported by the $200 \times 200$ pixels plot (figure \ref{fig:ds2}(b)) obtained by numerically integrating the field equations and characterizing their late-time states (black: $y = 1$ ; white $y^2 = \lambda$). As with $n = -1$, the viable phase space is the black region (described by a healthy on-shell limit $y = 1$). In the next section, it will further be shown that both de Sitter states are approached exponentially.

To end this section, we must mention that outside of the domains shown in figures \ref{fig:ds1} and \ref{fig:ds2} there is a region of phase space where the numerical integration fails, i.e., system becomes stiff after a few iterations and the demanded numerical precision cannot be achieved. In this region, the phase portraits can be used to confirm that the dynamical system eventually falls to $H_C\left[J\right]$ on which the vector $\left( x' , y' \right)$ blows up. We conjecture this as an artifact of the choice of coordinate system because Eq. \eqref{eq:conserved_current_j_st} casts no doubt that the cosmological dynamics should end up on the hypersurface $J \left( t \right) = 0$.

\subsection{Numerical experiments with multiple cosmic fluids}
\label{subsec:numerical_experiments}

We present numerical experiments with a nontrivial mixture of cosmic fluids. Starting from Eqs. \eqref{eq:friedmann_eq_gccg}, \eqref{eq:hubble_eq_gccg}, and \eqref{eq:sfe_gccg}, we add nonrelativistic matter ($w = 0$), radiation ($w = 1/3$), and vacuum energy ($w = -1$) to the cosmic fluid. In Eqs. \eqref{eq:friedmann_eq_gccg} and \eqref{eq:hubble_eq_gccg} this can be done by writing down
\begin{equation}
\rho = \dfrac{\rho_m}{a^3} + \dfrac{\rho_r}{a^4} + \rho_\Lambda
\end{equation}
and
\begin{equation}
\rho + P = \dfrac{\rho_m}{a^3} + \dfrac{4 \rho_r}{ 3 a^4 }
\end{equation}
where $a$ is the scale factor and $\rho_m$, $\rho_r$, and $\rho_\Lambda$ are constants corresponding to the amounts of nonrelativistic matter ($m$), radiation ($r$), and vacuum energy ($\Lambda$), respectively, that is available in the cosmic mixture. The Friedmann constraint, Hubble equation, and scalar field equation become
\begin{equation}
c a^3 \dot{\phi}^{2n} \left(h (2 n-1) a - 2 n \dot{a}\right)+h 2^n \left(a \left(-3 a \dot{a}^2+\rho_\Lambda  a^3+\rho_m\right)+\rho_r\right)=0 ,
\end{equation}
\begin{equation}
2 c n a^3 \dot{\phi}^{2n} \left(3 \dot{\phi} \left(h a - \dot{a}\right)+a \ddot{\phi}\right)+h 2^n \dot{\phi} \left(3 a \left(2 a^2 \ddot{a}-2 a \dot{a}^2+\rho_m\right)+4 \rho_r \right)=0 ,
\end{equation}
and
\begin{equation}
n \dot{\phi} \left(2 \dot{a}^2+a \left( \ddot{a}-3 h \dot{a}\right)\right)=n (2 n-1) a \ddot{\phi} \left(h a-\dot{a}\right) ,
\end{equation}
respectively. To numerically integrate with most convenience, as in the previous section, we nondimensionalize the variables using the constant $h$:
\begin{eqnarray}
\beta &=& c h^{2 (n-1)} \\
\tau &=& h t \\
a \left( \tau \right) &=& a \left( \tau /h \right) \\
\Psi \left( \tau \right) &=& \dot{\phi} \left( \tau/h \right) / h \\
\alpha_m &=& \rho_m / 3 h^2 \\
\alpha_r &=& \rho_r / 3 h^2 \\
\alpha_\Lambda &=& \rho_\Lambda / 3 h^2 .
\end{eqnarray}
In this way, we obtain the dimensionless version of the Friedmann constraint, Hubble equation, and scalar field equation given by
\begin{equation}
\label{eq:feq_gccg_nd}
3 \times 2^n \left(a \left(-a a'^2+\alpha_\Lambda a^3+\alpha_m\right)+\alpha_r\right)+\beta  a^3 \Psi^{2n} \left((2 n-1) a - 2 n a'\right)=0 ,
\end{equation}
\begin{equation}
\label{eq:heq_gccg_nd}
2 \beta  n a^3 \Psi^{2n} \left(a \left(\Psi'+3 \Psi \right)-3 \Psi a' \right)+3\ 2^n \Psi \left(2 a^3 a''-2 a^2 a'^2+3 \alpha_m a+4 \alpha_r\right)=0 ,
\end{equation}
and
\begin{equation}
\label{eq:seq_gccg_nd}
n \Psi \left(2 a'^2+a \left(a'' - 3 a' \right)\right)= n (2 n-1) a \left(a - a' \right) \Psi' ,
\end{equation}
respectively, where a prime denotes differentiation with respect to the dimensionless time $\tau$. The above system is described by the parameters $\{ n, \alpha_m, \alpha_r, \alpha_\Lambda, \beta \}$. Eqs. \eqref{eq:heq_gccg_nd} and \eqref{eq:seq_gccg_nd} should be integrated subject to the constraint Eq. \eqref{eq:feq_gccg_nd}. We note that the Friedmann constraint (Eq. \eqref{eq:feq_gccg_nd}) can be consider as a functional $\Psi \left[ a, a' \right]$ and so the phase space of the system is two-dimensional. In integration, after providing the initial conditions $\left( a \left( \tau_0 \right), a'\left(\tau_0 \right) \right)$, the scalar field $\Psi \left( \tau_0 \right)$ is determined using the Friedmann constraint to ensure that the integration begins precisely on the constraint surface. 

The dark energy equation of state $w_\phi = P_\phi / \rho_\phi$ can also be easily computed by identifying the dark energy density $\rho_\phi$ and pressure $P_\phi$ in the field equations. In the nondimensionalized set of variables above the dark energy equation of state, $w_\phi$, is given by
\begin{equation}
\label{eq:wde_nd}
w_\phi = - \dfrac{ 2 n \Psi '+3 \Psi }{ \Psi \left( 6 n H -6 n +3 \right)}
\end{equation}
where $H = a'/a$. It can easily be shown that on the de Sitter vacuum $\left( H = 1, \Psi' = 0 \right)$ this gives $w_\phi = -1$. On the other hand, for the unhealthy state $H = \sqrt{ \alpha_\Lambda }$ this gives $w_\phi = -1 / \left( 1 - 2n \right)$. These limits are confirmed by the numerical results.

\begin{figure}[h!]
\center
	\subfigure[ ]{
		\includegraphics[width = 0.48 \textwidth]{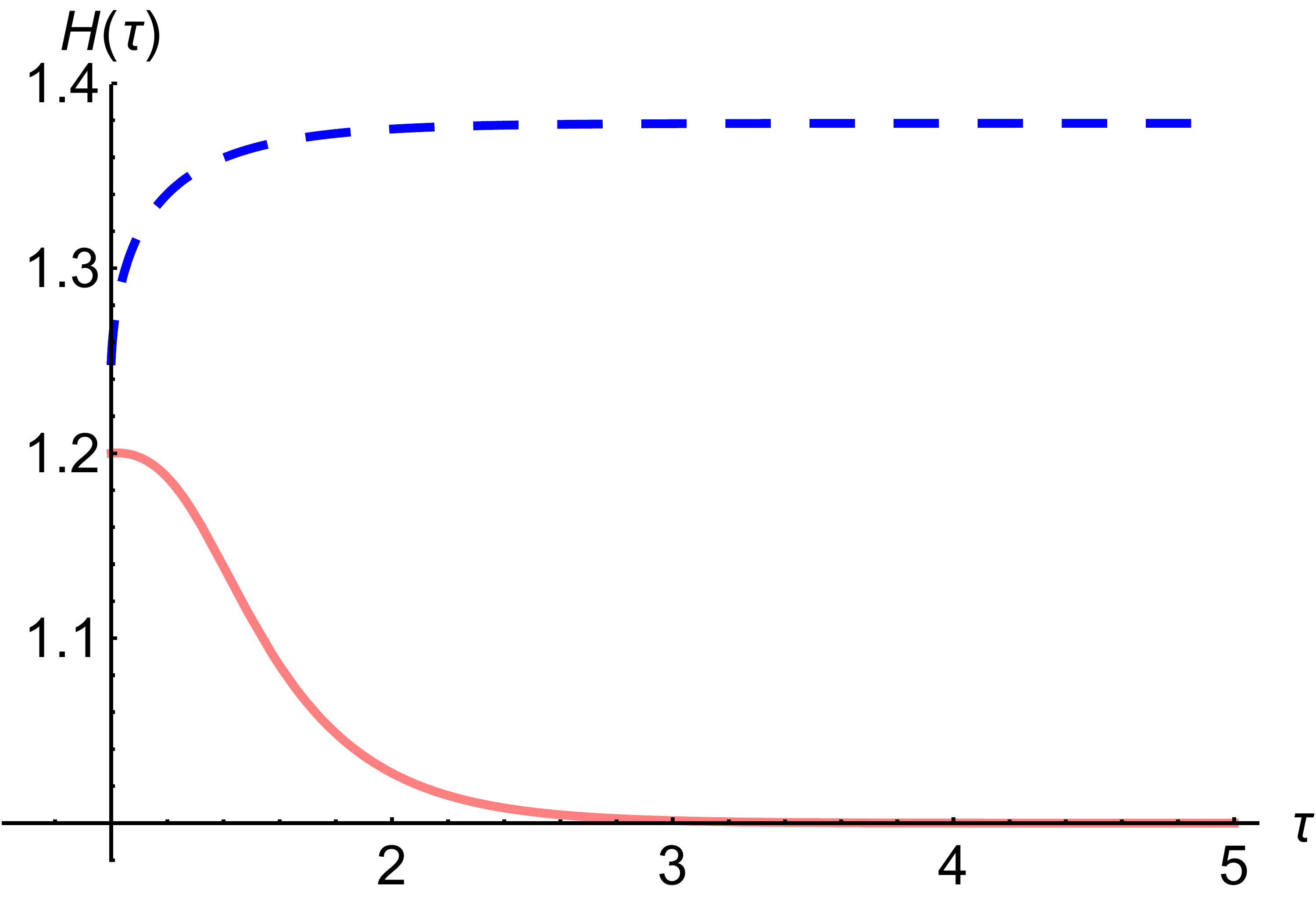}
		}
	\subfigure[ ]{
		\includegraphics[width = 0.48 \textwidth]{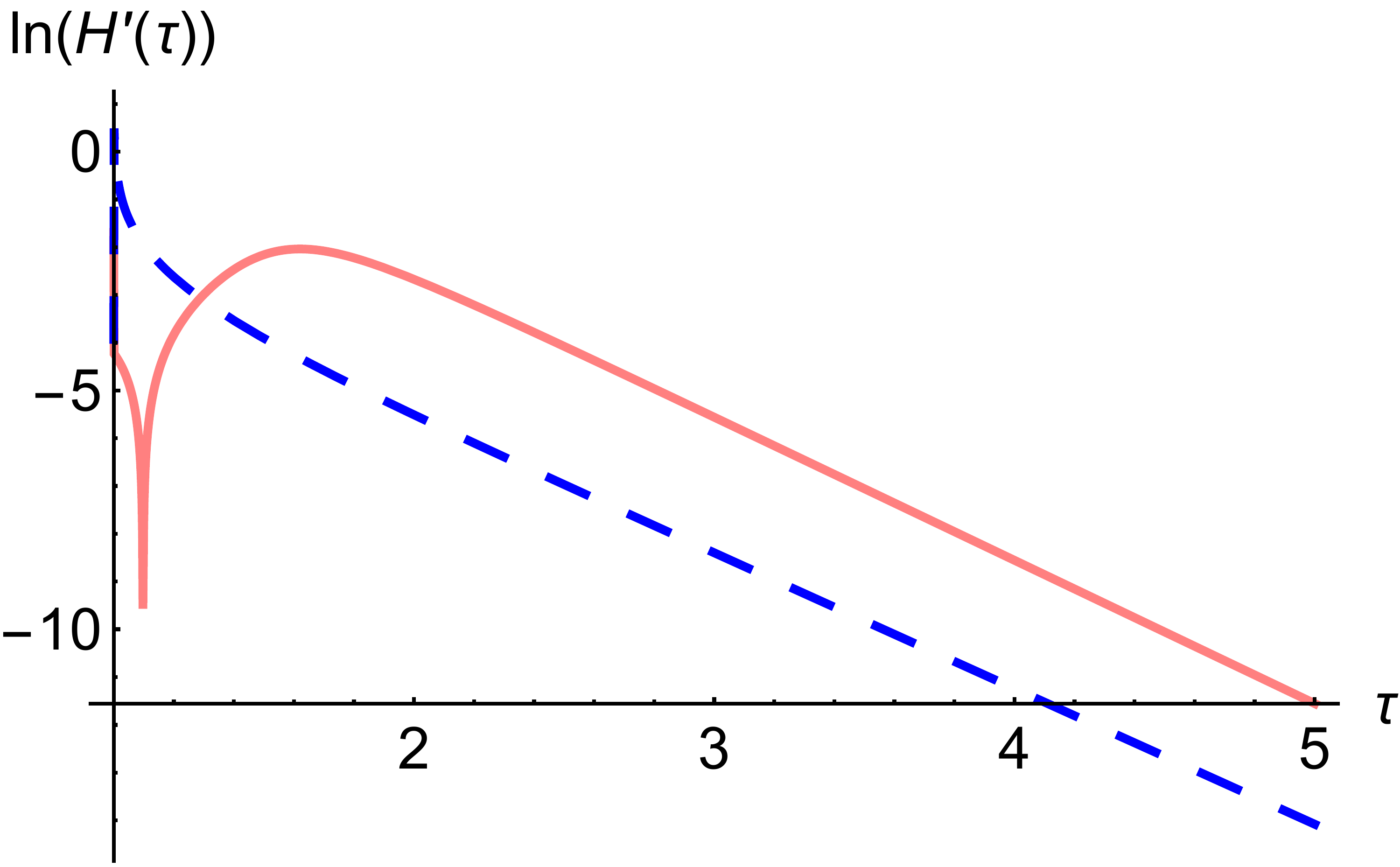}
		}
	\subfigure[ ]{
		\includegraphics[width = 0.48 \textwidth]{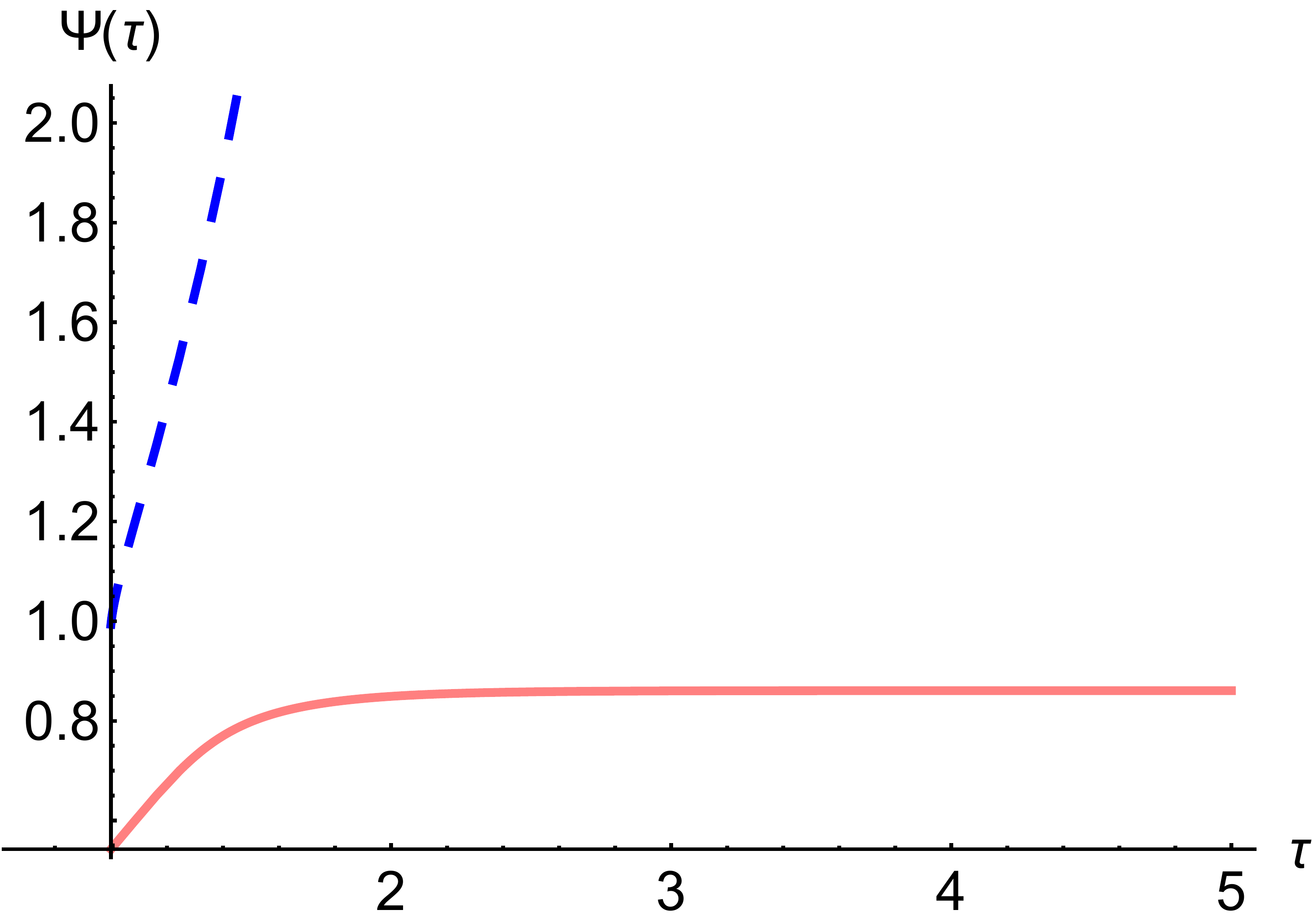}
		}
	\subfigure[ ]{
		\includegraphics[width = 0.48 \textwidth]{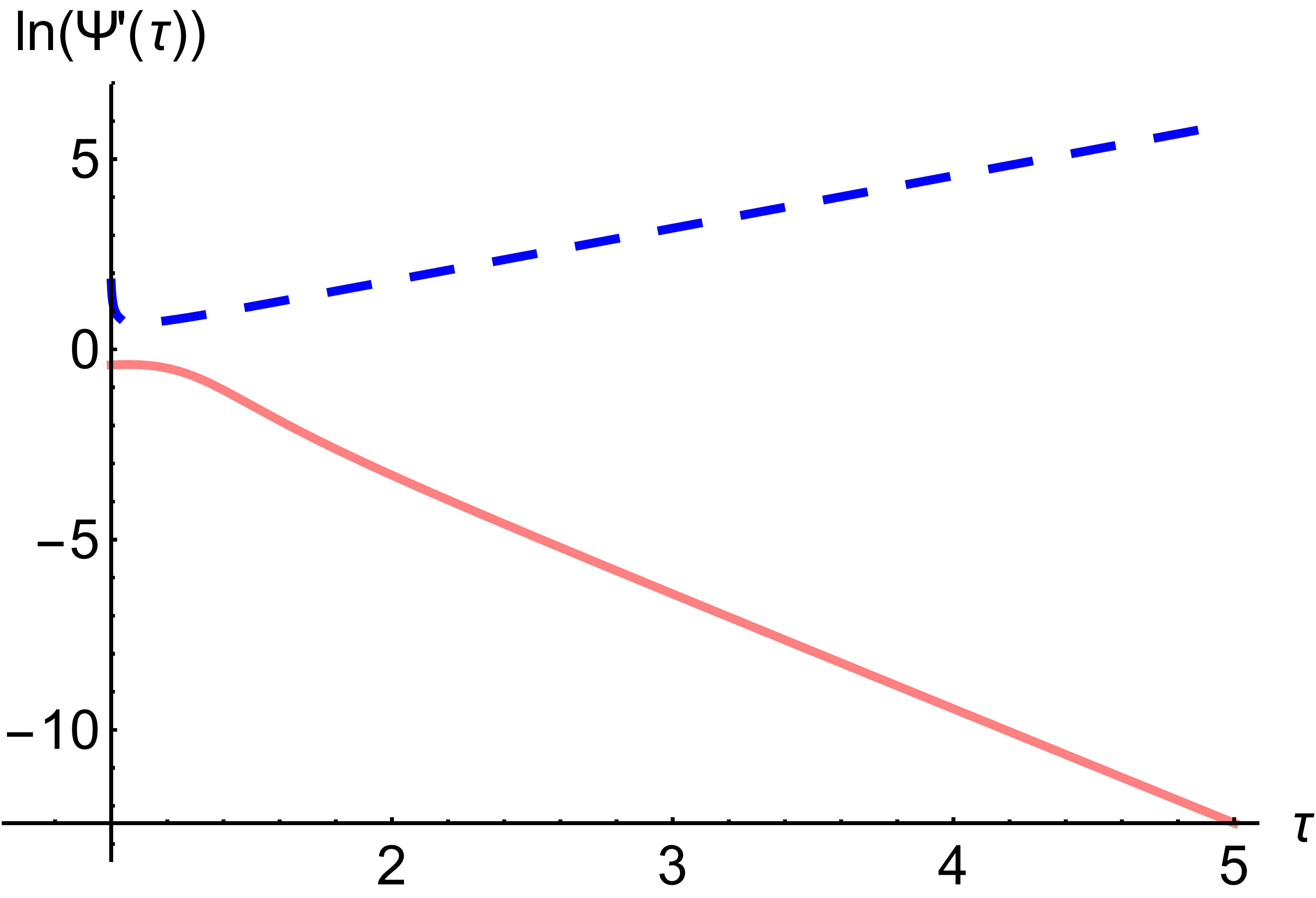}
		}
\caption{$\left[ n = -1 , \alpha_m = 8/10 , \alpha_r = 1/10, \alpha_\Lambda = 19/10, \beta = 1 \right]$ Plots of the (a) Hubble parameter, $H (\tau) = a'/a$, (b) $H' (\tau) = \left( a''/a \right) - \left( a'/a \right)^2$, (c) scalar field's velocity, $\Psi = \dot{\phi}/h$, and (d) scalar field's acceleration $\Psi'$ for $a'/a \left( \tau_0 \right) = 6/5$ (solid pink) and $a'/a\left( \tau_0 \right) = 5/4$ (dashed blue). The logarithms of $\Psi'$ and $H'$ reveal the exponential convergence towards the de Sitter vacuum.}
\label{fig:case1}
\end{figure}

\begin{figure}[h!]
\center
	\subfigure[ ]{
		\includegraphics[width = 0.48 \textwidth]{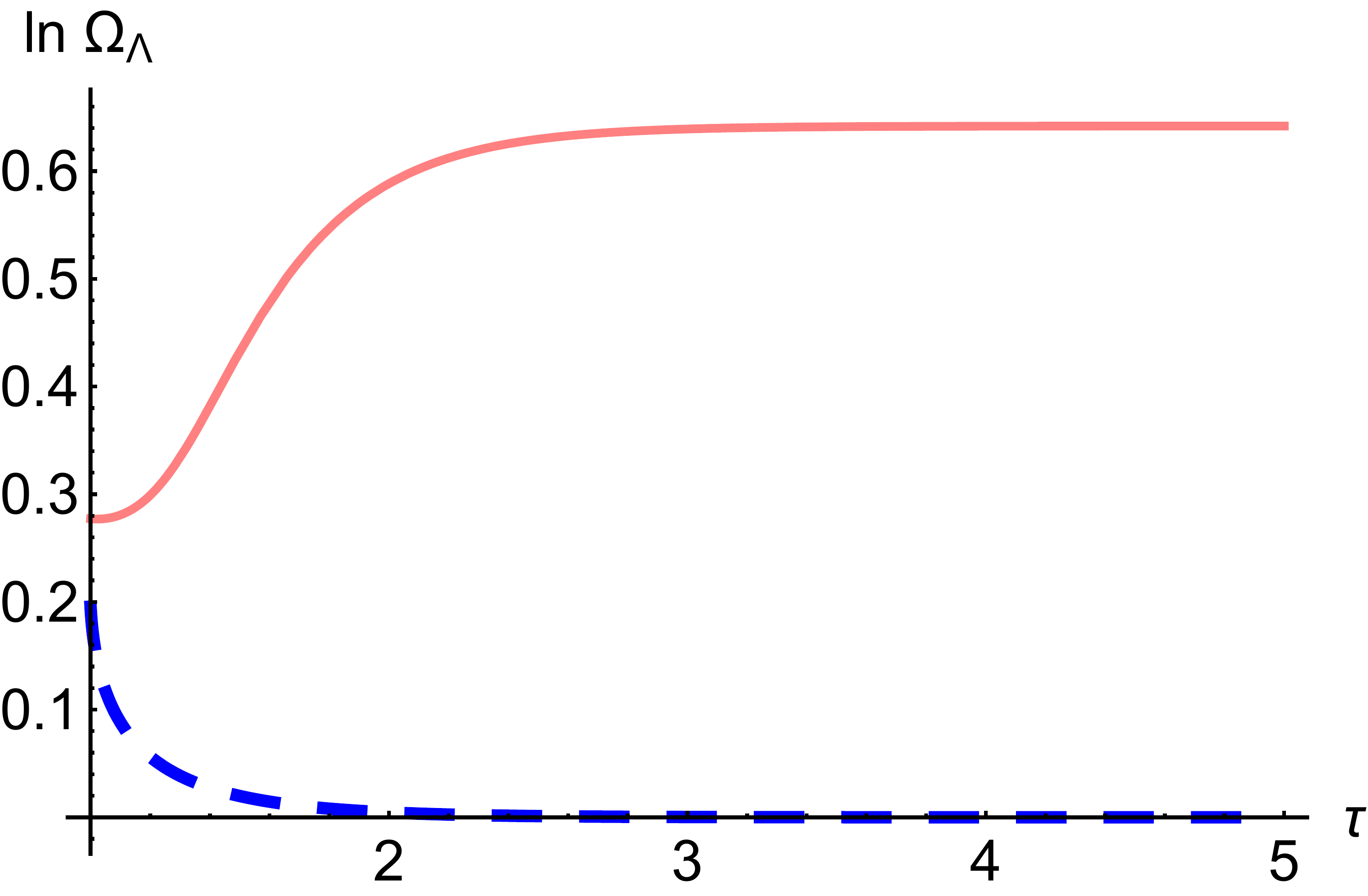}
		}
	\subfigure[ ]{
		\includegraphics[width = 0.48 \textwidth]{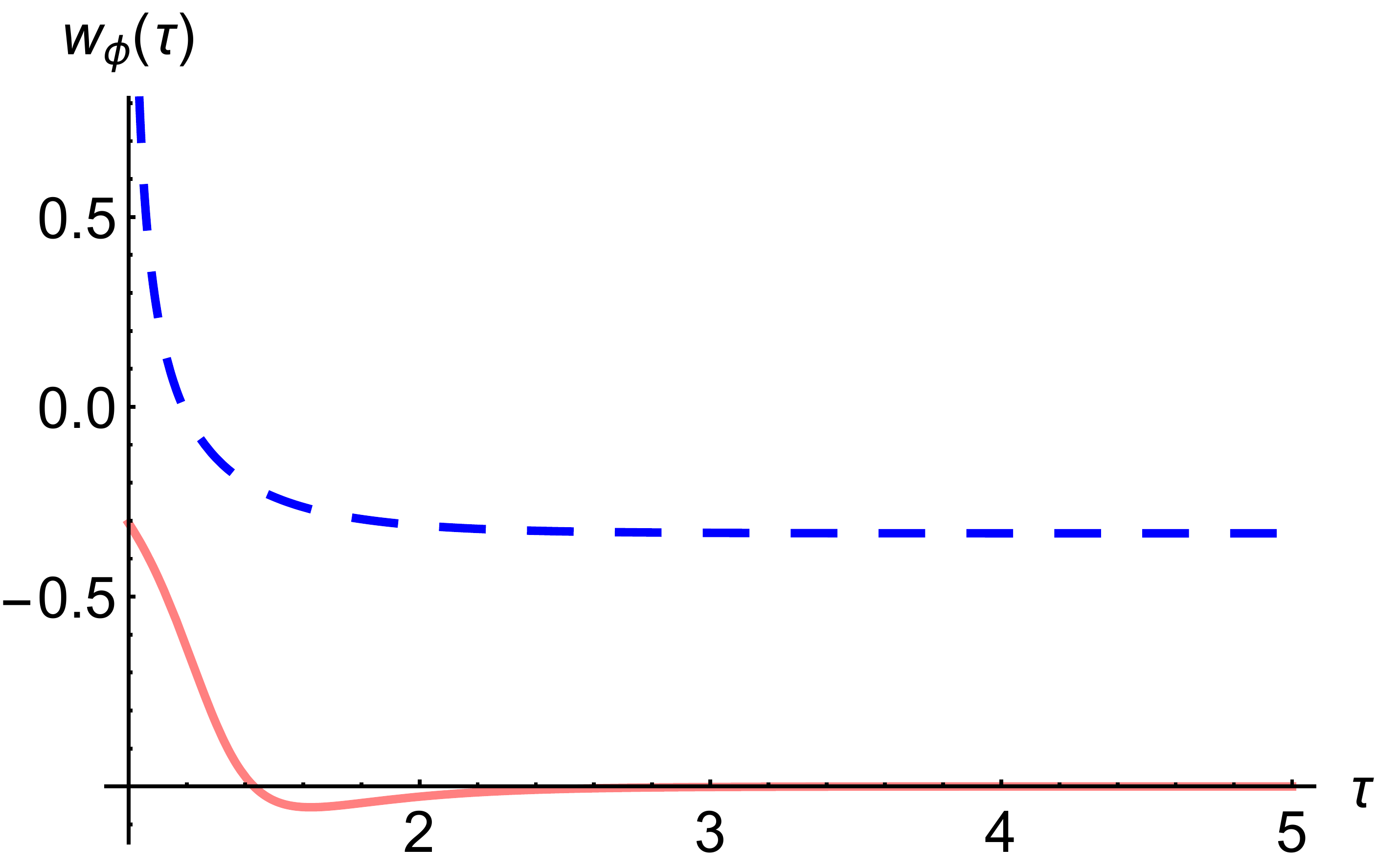}
		}
\caption{$\left[ n = -1 , \alpha_m = 8/10 , \alpha_r = 1/10, \alpha_\Lambda = 19/10, \beta = 1 \right]$ Evolution of (a) the vacuum energy density parameter, $\Omega_\Lambda = \rho_\Lambda/ \left( 3 H^2 \right)$, and (b) the dark energy equation of state, $w_\phi$, for $a'/a \left( \tau_0 \right) = 6/5$ (solid pink) and $a'/a\left( \tau_0 \right) = 5/4$ (dashed blue).}
\label{fig:case1vac}
\end{figure}

We present the numerical experiments for the same healthy theories studied in the previous section. For $\{ n = -1 , \alpha_m = 8/10 , \alpha_r = 1/10, \alpha_\Lambda = 19/10, \beta = 1 \}$, we obtain the results shown in figure \ref{fig:case1} for initial conditions $a'/a = 6/5$ (solid pink) and $a'/a = 5/4$ (dashed blue). 
Figures \ref{fig:case1}(a) and \ref{fig:case1}(b) confirm that there are indeed two de Sitter attractors (same $H = h$ and $3H^2 = \rho_\Lambda$ attractors of the previous section) and that these are approached exponentially. Also, figures \ref{fig:case1}(c) and \ref{fig:case1}(d) show that the the expected behavior of the scalar field on-shell (discussed in Sec. \ref{sec:self_tuned_kgb}) is recovered, i.e., on-shell $\dot{\phi} = q$ where $q$ is a constant. The on-shell limit of the scalar field is also clearly approached exponentially in \ref{fig:case1}(d). Figures \ref{fig:case1}(c) and \ref{fig:case1}(d) also show that the scalar field diverges on the de Sitter attractor $3H^2 = \rho_\Lambda$. The viable region of phase space is of course the two-dimensional region $\left( a \left( \tau_0 \right), a'\left(\tau_0 \right) \right)$ which goes to the healthy on-shell de Sitter vacuum, $H \left( \tau \right) = a' / a = 1$. Figure \ref{fig:case1vac} also shows how the density parameter of vacuum energy, $\Omega_\Lambda$, and the dark energy equation of state, $w_\phi$, evolve depending on the late-time state of the system. 
For states approaching the de Sitter vacuum, as the nonrelativistic and radiative cosmic fuels are used up, the system will eventually be left with only vacuum energy, $\rho_\Lambda$, and the dark scalar field's energy, $\rho_\phi$. This is clearly the case in figure \ref{fig:case1vac}(a) where the system settles down with $\Omega_\Lambda > 1$ (solid pink). On the other hand, the de Sitter state $3 H^2 = \rho_\Lambda$ is defined by a vanishing scalar field energy density and so the system is left only with vacuum energy at late times, $\Omega_\Lambda = 1$ (dashed blue in figure \ref{fig:case1vac}(a)). The dark energy equation of state strongly distinguishes the two de Sitter states (figure \ref{fig:case1vac}(b)). The de Sitter vacuum is supported by $w_\phi = -1$ whereas the unhealthy vacuum comes with $w_\phi = -1/ \left( 1 - 2n \right)$.

\begin{figure}[h!]
\center
	\subfigure[ ]{
		\includegraphics[width = 0.48 \textwidth]{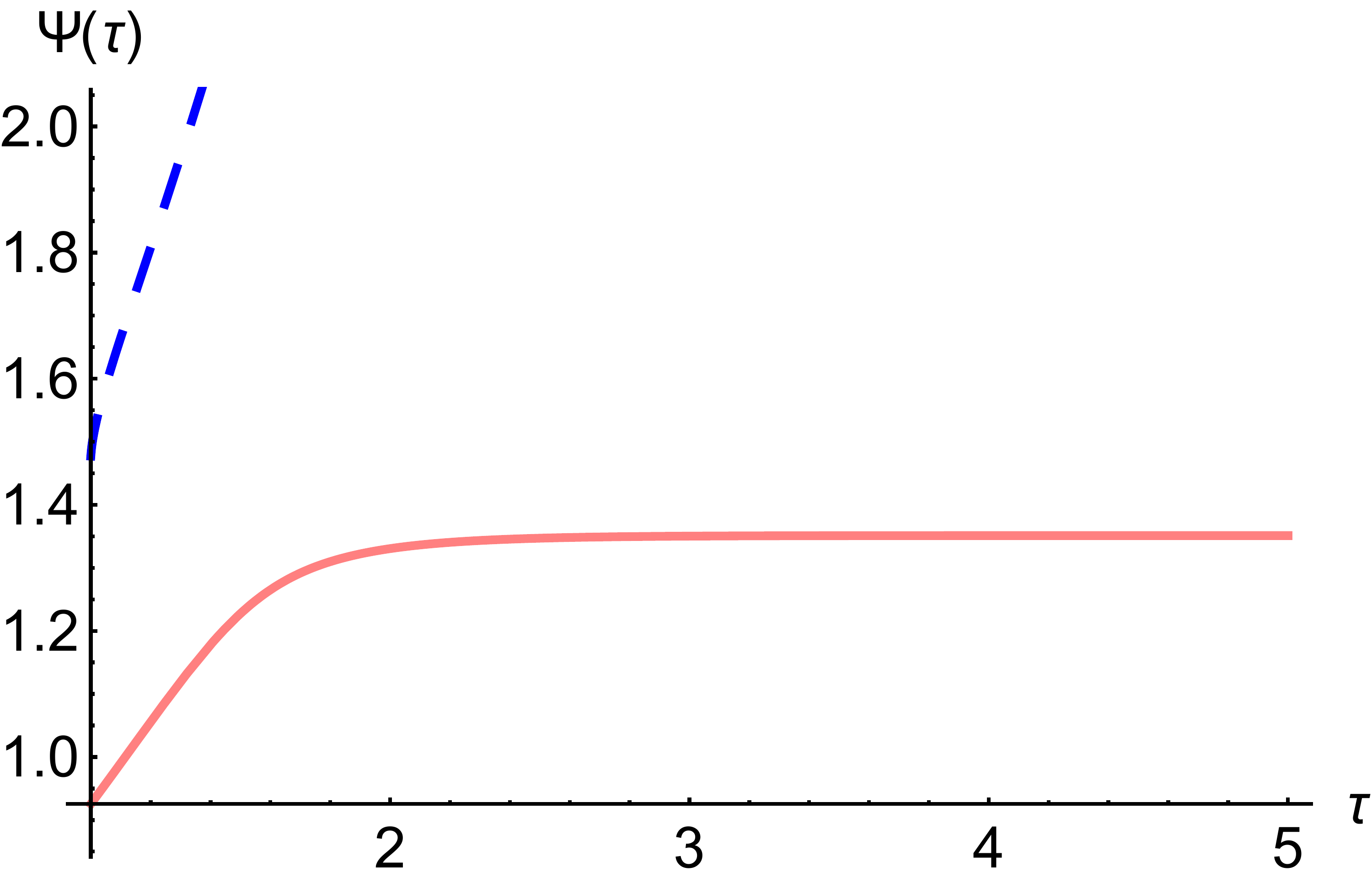}
		}
	\subfigure[ ]{
		\includegraphics[width = 0.48 \textwidth]{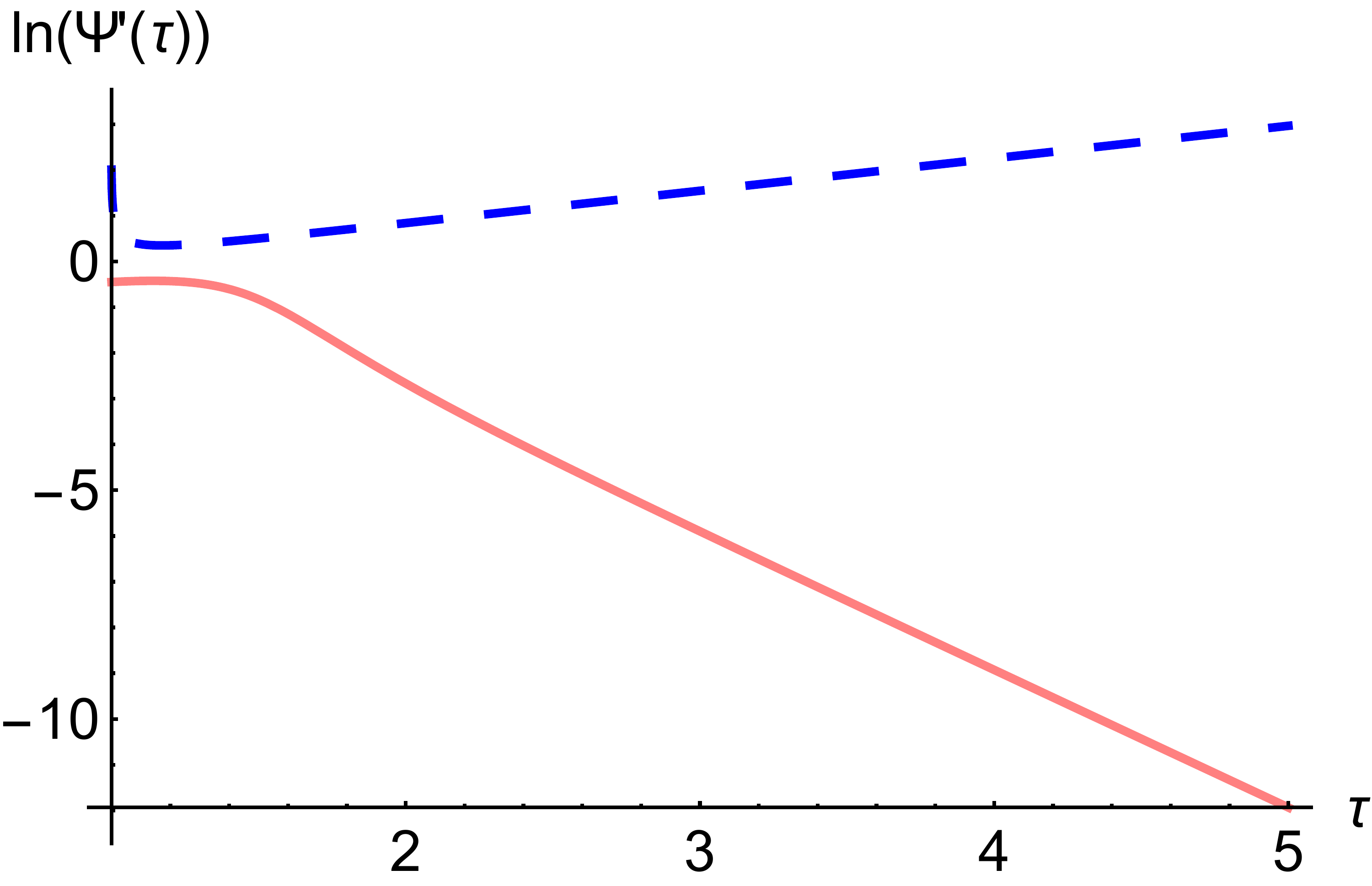}
		}
	\subfigure[ ]{
		\includegraphics[width = 0.48 \textwidth]{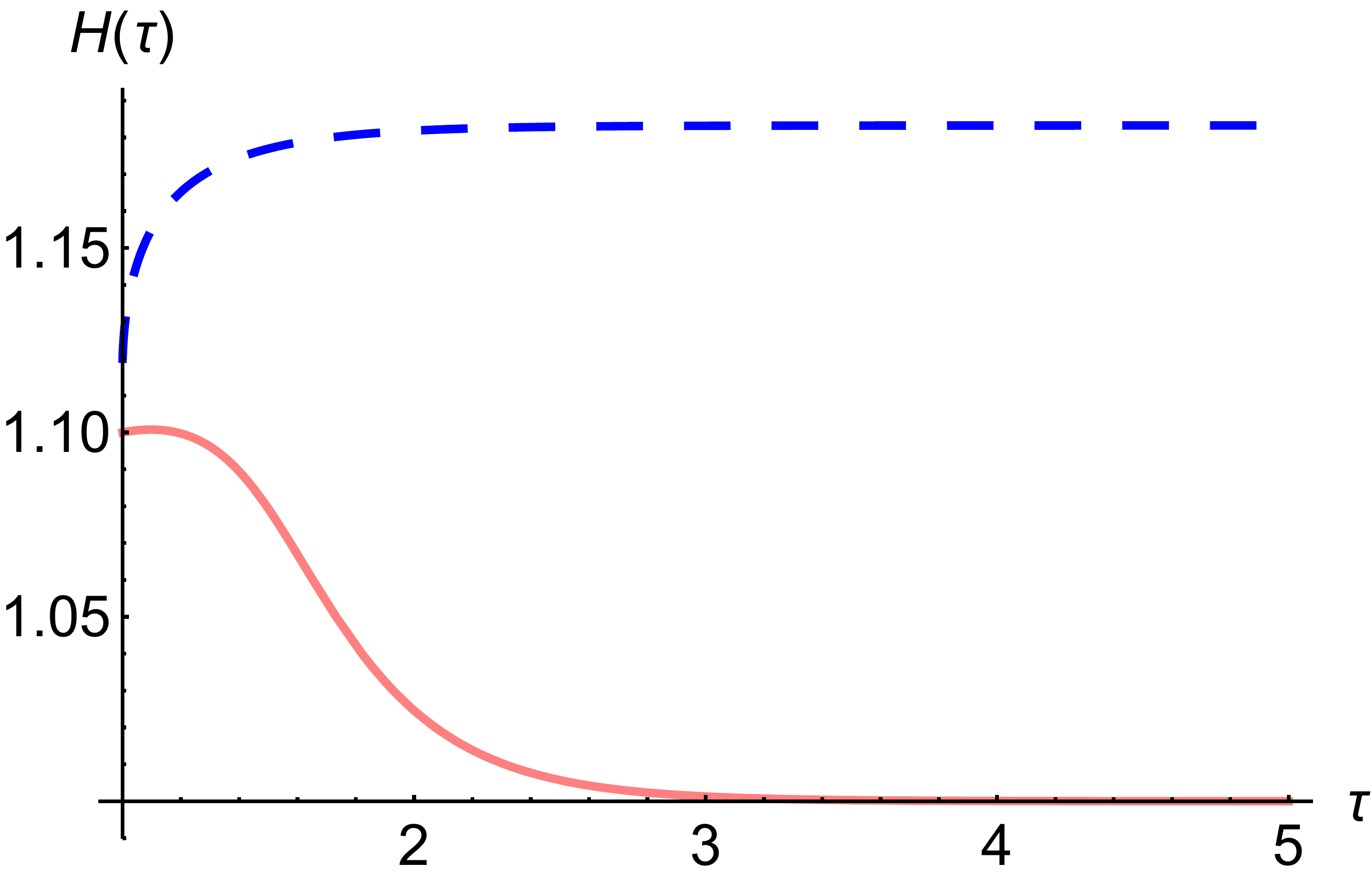}
		}
	\subfigure[ ]{
		\includegraphics[width = 0.48 \textwidth]{case2dh.pdf}
		}
\caption{$\left[ n = -2 , \alpha_m = 8/10 , \alpha_r = 1/10, \alpha_\Lambda = 7/5, \beta = 1 \right]$ Plots of the (a) Hubble parameter, $H (\tau) = a'/a$, (b) $H' (\tau) = \left( a''/a \right) - \left( a'/a \right)^2$, (c) scalar field's velocity, $\Psi = \dot{\phi}/h$, and (d) scalar field's acceleration $\Psi'$ for $a'/a \left( \tau_0 \right) = 11/10$ (solid pink) and $a'/a\left( \tau_0 \right) = 56/55$ (dashed blue). The logarithms of $\Psi'$ and $H'$ reveal the exponential convergence towards the de Sitter vacuum.}
\label{fig:case2}
\end{figure}

\begin{figure}[h!]
\center
	\subfigure[ ]{
		\includegraphics[width = 0.48 \textwidth]{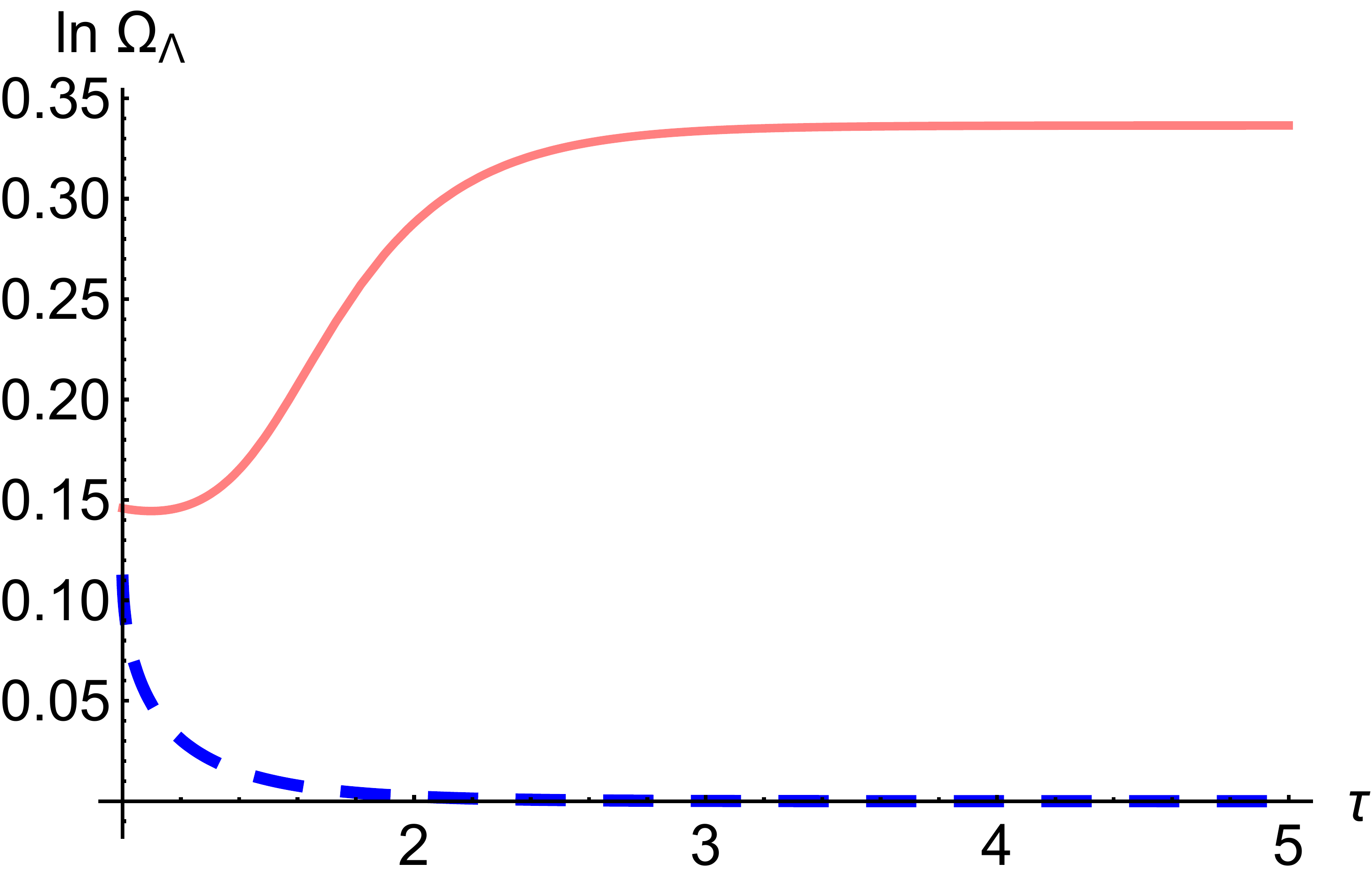}
		}
	\subfigure[ ]{
		\includegraphics[width = 0.48 \textwidth]{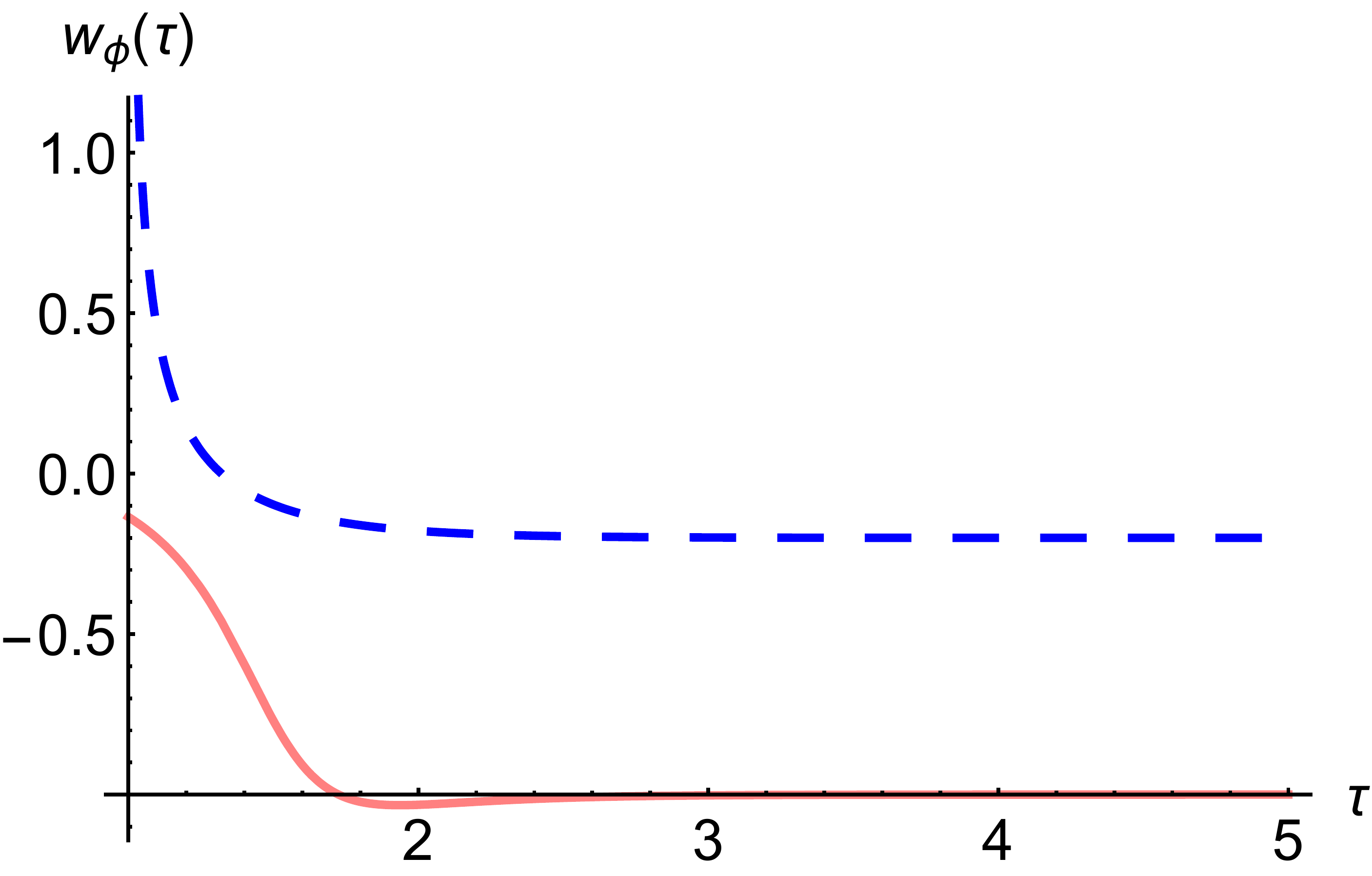}
		}
\caption{$\left[ n = -2 , \alpha_m = 8/10 , \alpha_r = 1/10, \alpha_\Lambda = 7/5, \beta = 1 \right]$ Evolution of (a) the vacuum energy density parameter, $\Omega_\Lambda = \rho_\Lambda/ \left( 3 H^2 \right)$, and (b) the dark energy equation of state, $w_\phi$, for $a'/a \left( \tau_0 \right) = 11/10$ (solid pink) and $a'/a\left( \tau_0 \right) = 56/55$ (dashed blue).}
\label{fig:case2vac}
\end{figure}

Similar conclusions can be obtained for $\{ n = -2 , \alpha_m = 8/10 , \alpha_r = 1/10, \alpha_\Lambda = 7/5, \beta = 1 \}$ as shown in figures \ref{fig:case2} and \ref{fig:case2vac} for initial conditions $a'/a = 11/10$ (solid pink) and $a'/a = 56/55$ (dashed blue). 
Again, the results show that the system converges exponentially to both de Sitter states. The on-shell de Sitter state is approached with a constant scalar field velocity and the unhealthy state with a diverging scalar field. Also, figure \ref{fig:case2vac} confirms the interpretation that the on-shell vacuum is supported by a constant scalar field velocity and $w_\phi = -1$ while the unhealthy de Sitter limit has a vanishing scalar field energy density and $w_\phi = -1 / \left( 1 - 2n \right)$.

\subsection{Stability through a phase transition}
\label{subsec:phase_transition}

It was shown in Refs. \cite{st_horndeski_cubic_appleby, st_horndeski_cosmology_emond2018} that the tempered de Sitter vacuum remains stable even under a phase transition. For the self-tuning generalized cubic covariant Galileon theory, however, the de Sitter vacuum cannot be stable for arbitrarily large phase transitions because of the existence of the attractor $3H^2 = \rho_\Lambda$ (see figures \ref{fig:ds1} and \ref{fig:ds2}). We present the stable phase transitions in this section.

\begin{figure}[h!]
\center
	\subfigure[ ]{
		\includegraphics[width = 0.48 \textwidth]{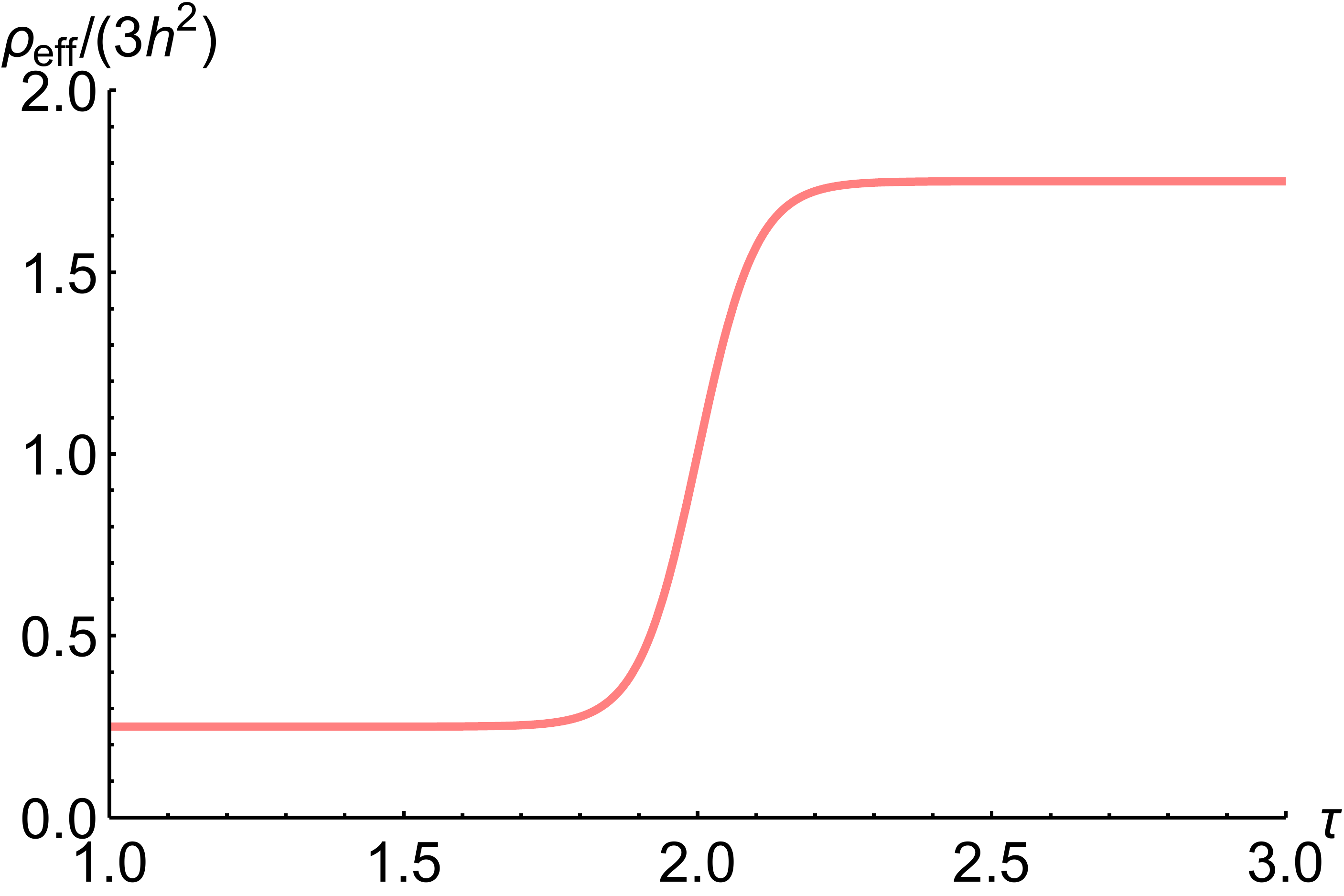}
		}
	\subfigure[ ]{
		\includegraphics[width = 0.48 \textwidth]{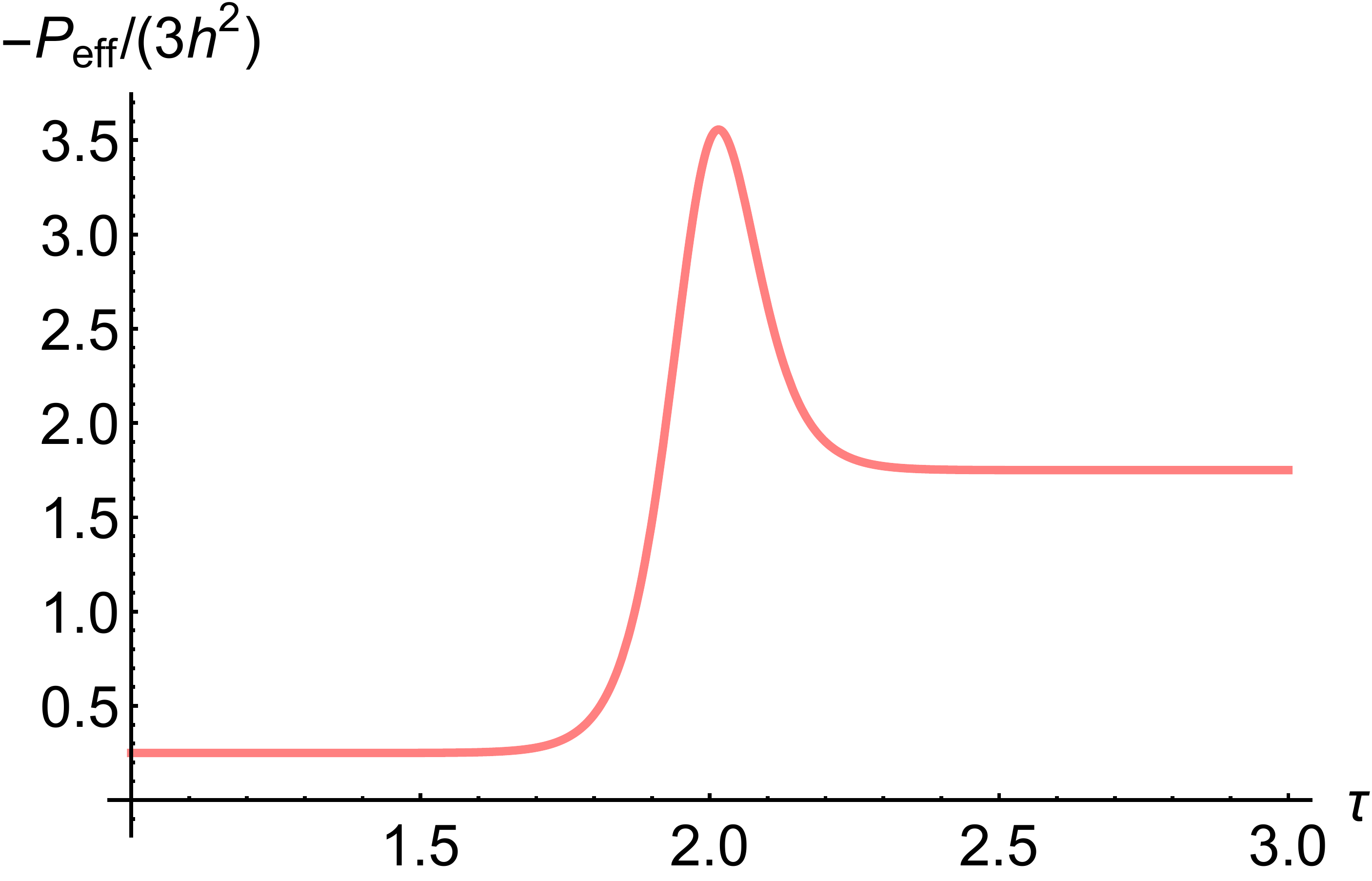}
		}
\caption{Effective energy density $\rho_{\text{eff}}$ and pressure $P_{\text{eff}}$ for phase transition experiment.}
\label{fig:pt_source}
\end{figure}

To implement the phase transition, we consider the effective energy density and pressure given by \cite{st_horndeski_cubic_appleby, st_horndeski_cosmology_emond2018}
\begin{equation}
\label{eq:rho_eff}
\rho_{\text{eff}} = \rho_\Lambda + \dfrac{ \Delta \rho_\Lambda }{2} \tanh \left( \dfrac{t - T}{\Delta t} \right)
\end{equation}
and
\begin{equation}
\label{eq:P_eff}
P = - \rho - \dfrac{ \dot{\rho} }{ 3H } ,
\end{equation}
respectively, which describe a fluid with vacuum energy transitioning from $\rho_\Lambda - \Delta \rho_\Lambda/2$ for $t \ll T$ to $\rho + \Delta \rho_\Lambda/2$ for $t \gg T$ during a short interval $\Delta t$ at time $t = T$. This behavior is reflected in figure \ref{fig:pt_source} where $\tau = h t$, $h T = 2$, $\Delta \tau = h \Delta T$, $\rho_\Lambda/ 3h^2 = 1$, and $\Delta \rho_\Lambda / 3h^2 = 3/2$. Clearly, the energy density transitions from $\rho_\Lambda - \Delta \rho_\Lambda / 2 = \left( 3 h^2 \right) /4$ to $\rho_\Lambda + \Delta \rho_\Lambda / 2 = \left( 3 h^2 \right) 7/4$ in the short time $h \Delta \tau = 10^{-1}$ at $h T = 2$. The negative pressure, $-P_{\text{eff}}$, on the other hand, transitions from $\rho_\Lambda - \Delta \rho_\Lambda/2$ to $\rho_\Lambda + \Delta \rho_\Lambda/2$ in the same time scale but in the top hat manner as shown in figure \ref{fig:pt_source}. Far away from the transition ($t \ll T$ and $t \gg T$) we see that the fluid acts like vacuum energy, i.e., $w = -1$.

Starting from the field equations, we setup the dynamical system first by eliminating $\rho$ using the Friedmann constraint (Eq. \eqref{eq:friedmann_eq_gccg}) and the Hubble equation (Eq. \eqref{eq:hubble_eq_gccg}). This leads to the pressure equation which we integrate together with the scalar field equation for the vector $\left( \dot{\phi}, H \right)$. Nondimensionalizing the variables using the following transformations
\begin{eqnarray}
\beta &=& c h^{2 (n-1)} \\
\tau &=& h t \\
\Delta \tau &=& h \Delta t \\
\mathcal{T} &=& h T \\
y \left( \tau \right) &=& H \left( \tau /h \right) / h \\
\Psi \left( \tau \right) &=& \dot{\phi} \left( \tau/h \right) / h \\
\lambda &=& \rho_\Lambda / 3 h^2 \\
\Delta \lambda &=& \Delta \rho_\Lambda / 3 h^2
\end{eqnarray}
we obtain the dynamical system for $\left( \Psi, y \right)$ given by
\begin{equation}
\label{eq:ds1_pt}
\begin{split}
\Psi '= \dfrac{ 3 }{  \mathcal{D} \left( \tau \right)  } \bigg[ 2 \beta  \Delta \tau y \Psi ^{2n + 1} 
& -2^n \Psi (\tau ) \bigg(\Delta \lambda  \text{sech}^2\left(\frac{\tau -T}{\Delta \tau }\right) \\
& +3 \Delta \tau  y \left(\Delta \lambda  \tanh \left(\frac{\tau -T}{\Delta \tau }\right)+2 \lambda +2 (y-2) y\right)\bigg) \bigg]
\end{split}
\end{equation}
and
\begin{equation}
\label{eq:ds2_pt}
\begin{split}
y' = \dfrac{ 3 (y-1) }{\mathcal{D} \left( \tau \right)} \bigg[ 2 \beta  \Delta \tau  y \Psi ^{2n} (2 n y-2 n+1)
&+2^n (2 n-1) \bigg( \Delta \lambda  \text{sech}^2\left(\frac{\tau -T}{\Delta \tau }\right) \\
&+3 \Delta \tau  y \left(\Delta \lambda  \tanh \left(\frac{\tau -T}{\Delta \tau }\right)+2 \lambda -2 y^2\right) \bigg) \bigg]
\end{split}
\end{equation}
where a prime denotes differentiation with respect to $\tau$ and $\mathcal{D} \left( \tau \right)$ is given by
\begin{equation}
\mathcal{D} \left( \tau \right) = 4 \Delta \tau  y \left(3 \times 2^n (2 n-1) (y-1)-\beta  n \Psi^{2n}\right) .
\end{equation}
The dynamical variables $\left( \Psi, y \right)$ can be solved from Eqs. \eqref{eq:ds1_pt} and \eqref{eq:ds2_pt} upon providing the initial conditions $\left( \Psi \left( \tau_0 \right), y \left( \tau_0 \right) \right)$ at $\tau = \tau_0$. In the following examples, we choose the initial conditions such that the no-ghost constraint is satisfied before and after the phase transition. 

\begin{figure}[h!]
\center
	\subfigure[ ]{
		\includegraphics[width = 0.48 \textwidth]{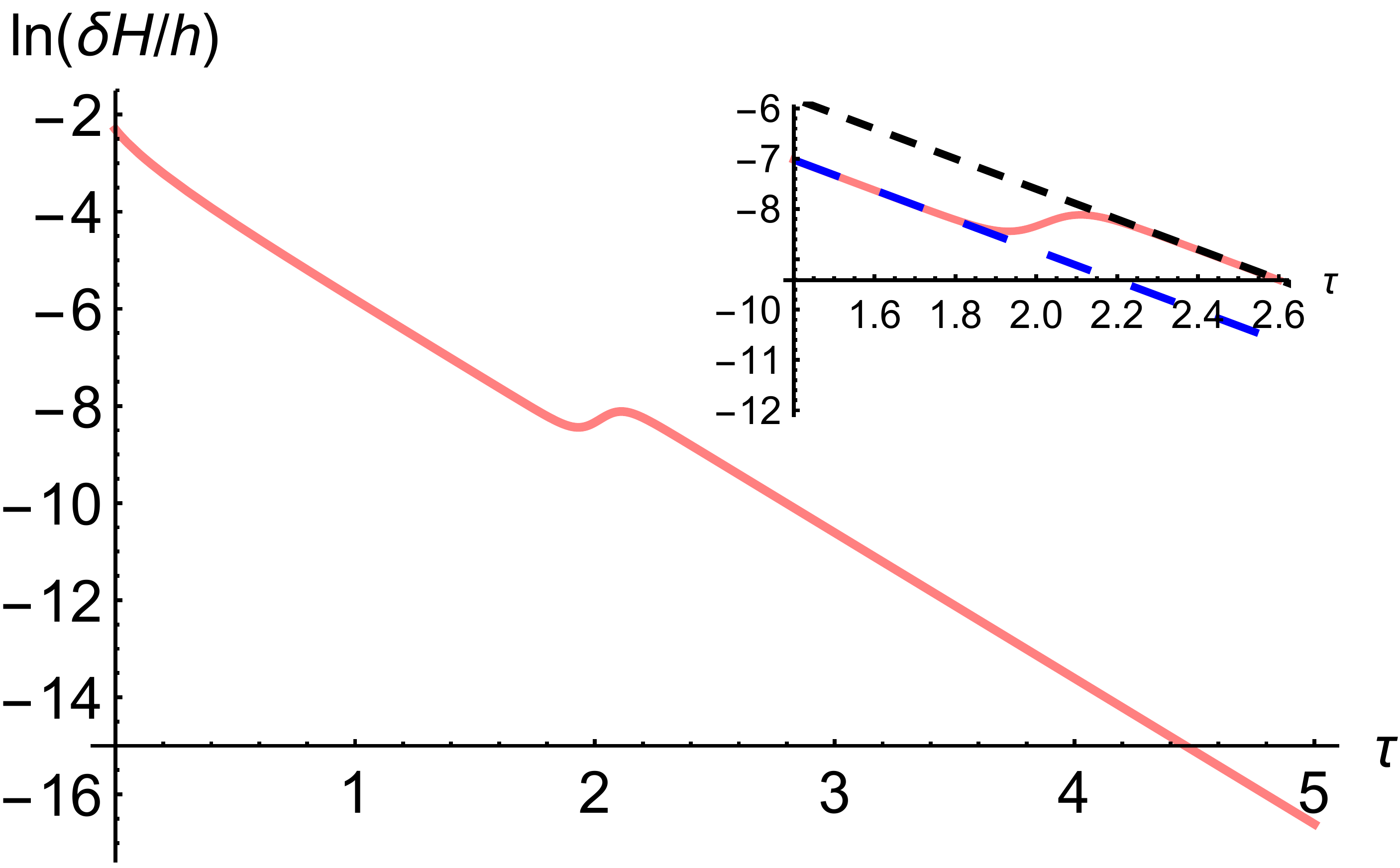}
		}
	\subfigure[ ]{
		\includegraphics[width = 0.48 \textwidth]{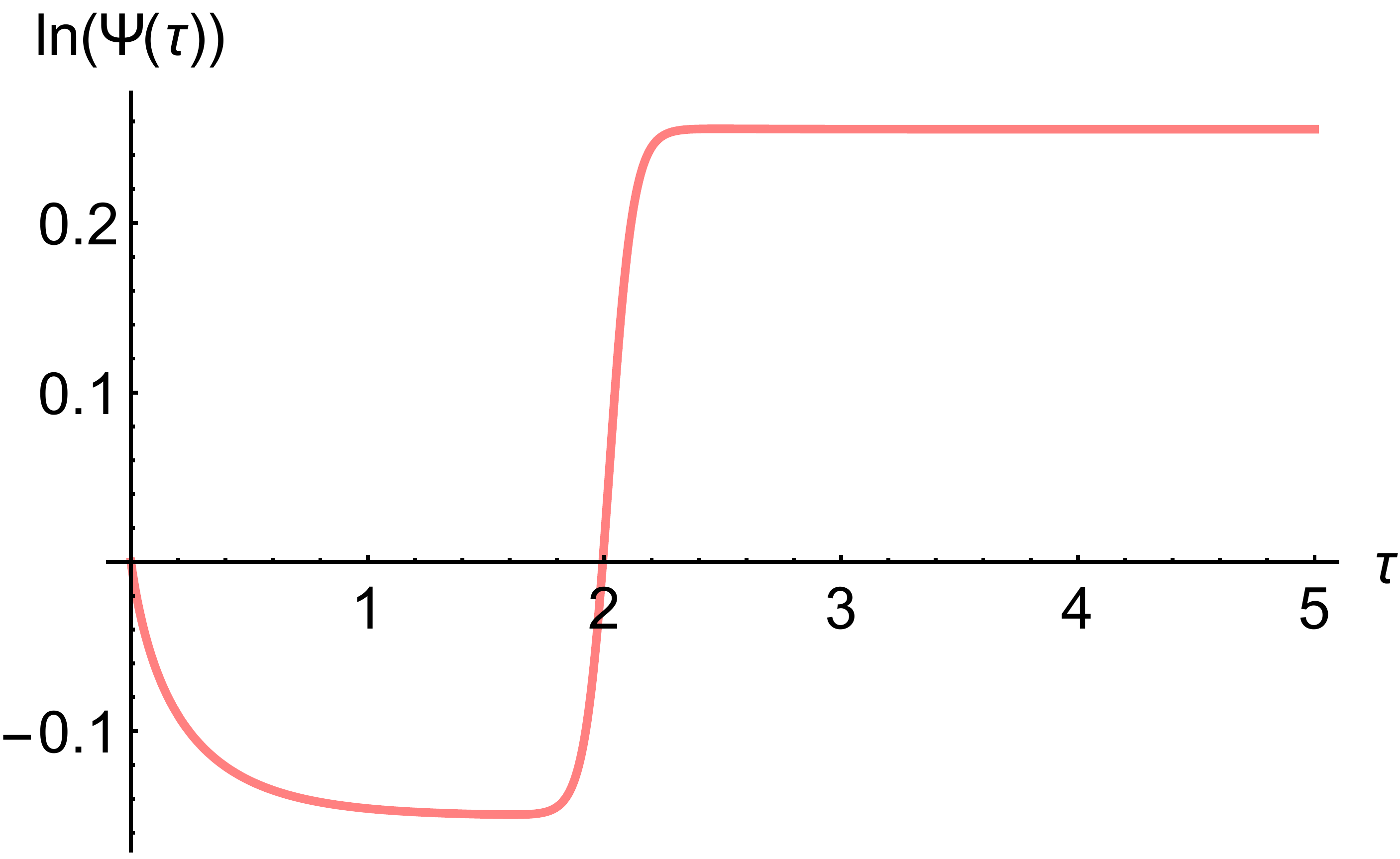}
		}
\caption{$\left[ n = -1, \beta = 1 \right]$ Logarithmic plots of (a) the change in the Hubble parameter, $\ln \left( \left( H/ h \right) - 1 \right)$, and (b) the scalar field's velocity, $\ln \left(\Psi\right) = \ln\left( \dot{\phi}/h \right)$, as the system undergoes a phase transition from $\lambda - \Delta \lambda/2= 19/10$ to $\lambda + \Delta \lambda/2 = 7/5$ for a time interval $\delta \tau = 10^{-1}$ at $\tau = 2$. The inset plot to (a) shows the asymptotic solution $\delta H/h \sim e^{-3 \tau}$ in the phases $\lambda - \Delta \lambda / 2$ (long dash blue) and $\lambda + \Delta \lambda / 2$ (short dash black).}
\label{fig:pt1}
\end{figure}

\begin{figure}[h!]
\center
	\subfigure[ ]{
		\includegraphics[width = 0.48 \textwidth]{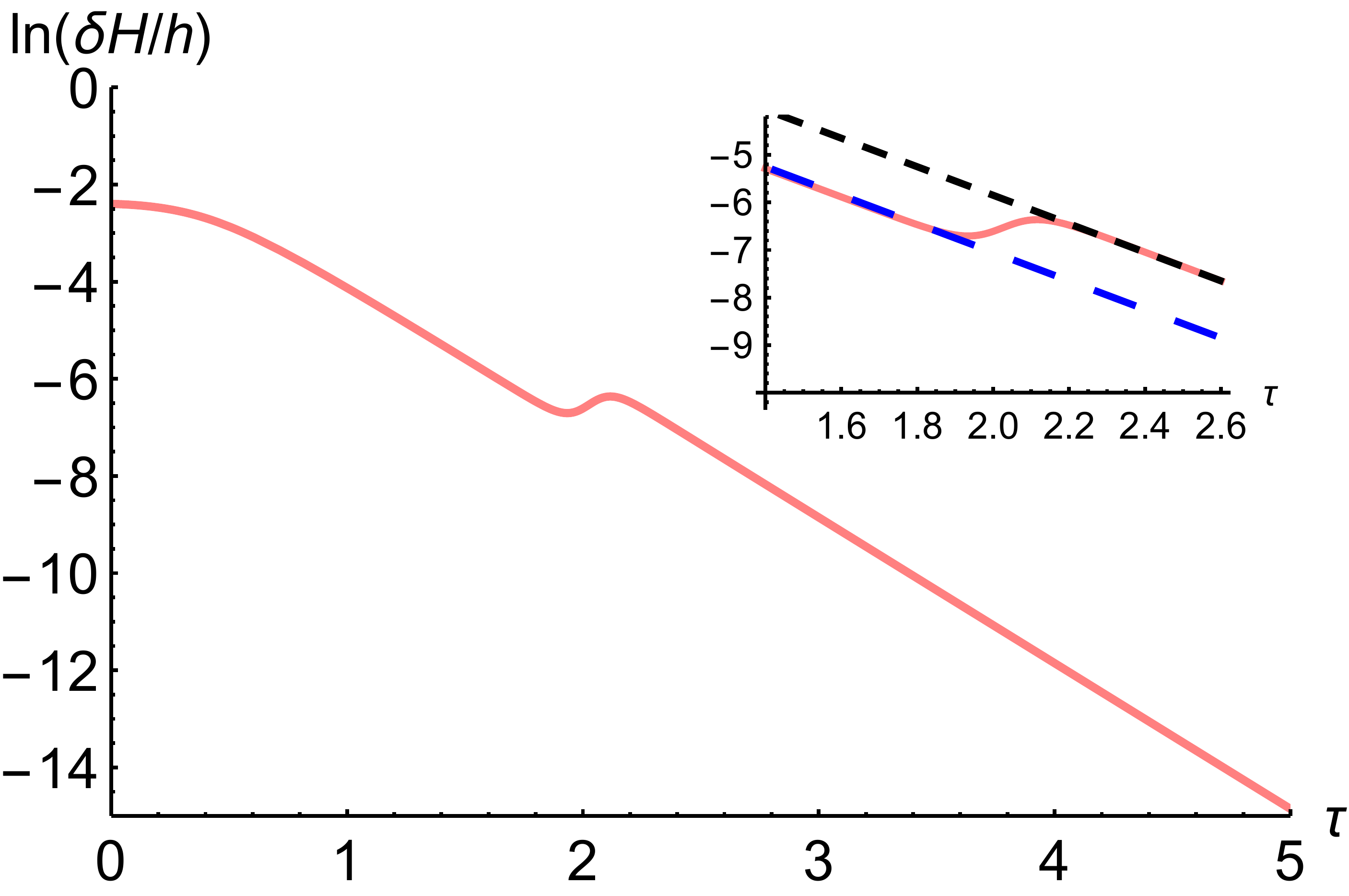}
		}
	\subfigure[ ]{
		\includegraphics[width = 0.48 \textwidth]{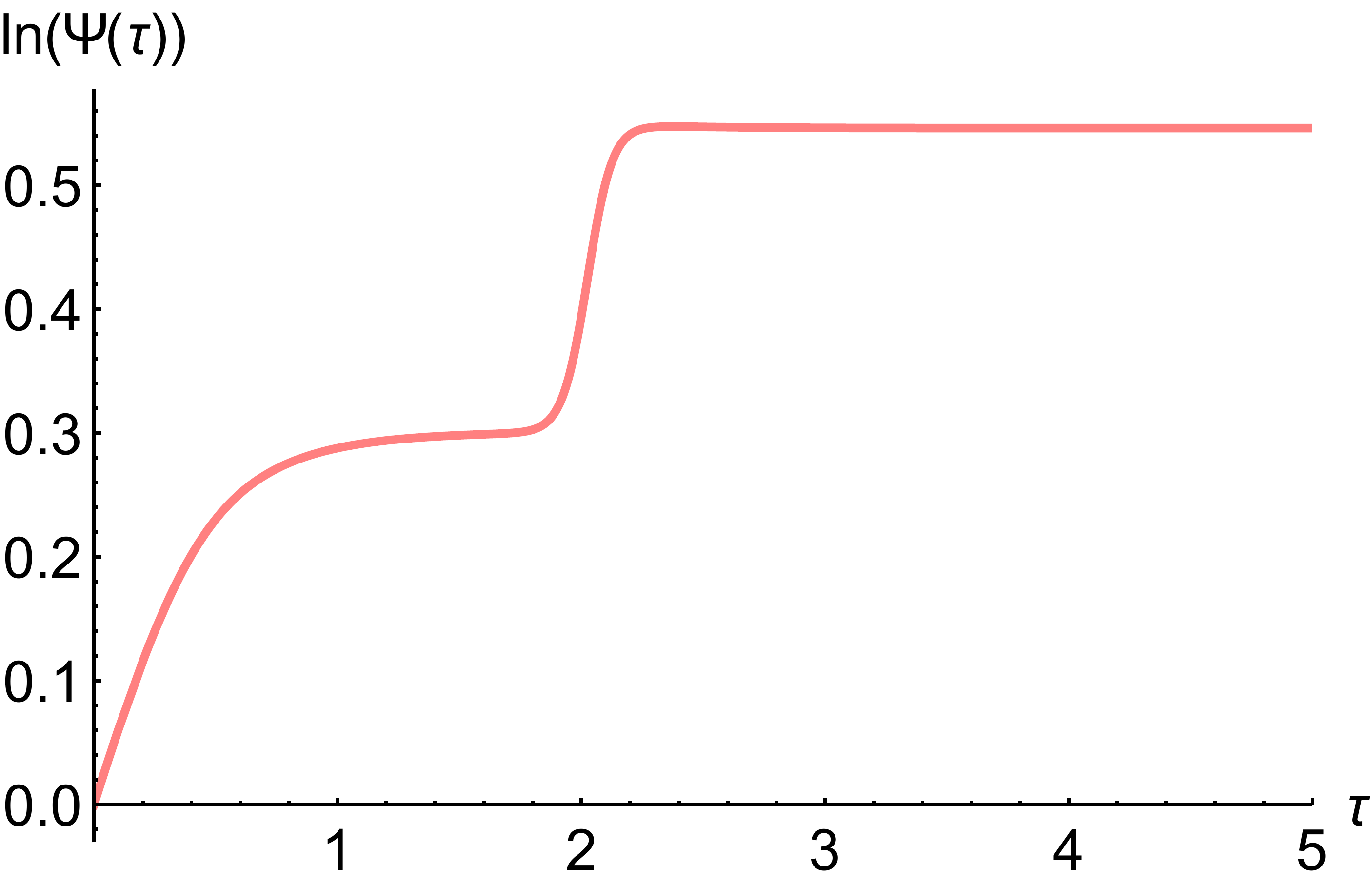}
		}
\caption{$\left[ n = -2, \beta = 1 \right]$ Logarithmic plots of (a) the change in the Hubble parameter, $\ln \left( \left( H/ h \right) - 1 \right)$, and (b) the scalar field's velocity, $\ln \left(\Psi\right) = \ln\left( \dot{\phi}/h \right)$, as the system undergoes a phase transition from $\lambda - \Delta \lambda / 2 = 7/5$ to $\lambda + \Delta \lambda / 2 = 23/20$ for a time interval $\delta \tau = 10^{-1}$ at $\tau = 2$. The inset plot to (a) shows the asymptotic solution $\delta H/h \sim e^{-3 \tau}$ in the phases $\lambda - \Delta \lambda/2$ (long dash blue) and $\lambda + \Delta \lambda / 2$ (short dash black).}
\label{fig:pt2}
\end{figure}

Figure \ref{fig:pt1} shows the Hubble parameter and the scalar field velocity responding to the phase transition of interval $\Delta \tau = 10^{-1}$ at $\tau = 2$ for the case $n = -1$ and $\beta = 1$.
In this case, the initial state with $\lambda - \Delta \lambda/2= 19/10$ was chosen to correspond to the same cases studied in Secs. \ref{subsec:dynamical_system_analysis} and \ref{subsec:numerical_experiments} for $n = -1$ while $\Delta \lambda = -1/2$ was arbitrarily chosen with the only condition that the no-ghost constraint must continue to be satisfied in the final state. As expected from the stable vacuum, the Hubble parameter (figure \ref{fig:pt1}(a)) continues to exponentially converge to $H = h$ even after the phase transition while at the same time the scalar field's velocity (figure \ref{fig:pt1}(b)) receives a constant shift. The trajectory change of the Hubble parameter is also revealed in detail near the phase transition by plotting the numerical solution together with the asymptotic solution $\delta H \sim e^{-3 h t}$ (Eq. \eqref{eq:h_asymp}) in both phases as shown in the inset of figure \ref{fig:pt1}(a). Figure \ref{fig:pt2} shows the corresponding results for $n = -2$ and $\beta = 1$. The initial state was chosen to match the same cases studied in Secs. \ref{subsec:dynamical_system_analysis} and \ref{subsec:numerical_experiments} for $n = -2$ while $\Delta \lambda = -1/4$ was arbitrarily chosen as long as the final state is healthy. 
Stability is again revealed as the Hubble parameter transitions from the initial state to the final state (figure \ref{fig:pt2}) as shown by the asymptotic trajectories $\delta H \sim e^{-3 h t}$ while the scalar field's velocity shifts by a constant value.

\subsection{Can a self-tuning GCCG screen a large bare cosmological constant?}
\label{subsec:self_tuning_gccg_vs_large_cc}

In the previous subsections, we have shown using numerical examples that the self-tuning generalized cubic covariant Galileon (GCCG) do indeed possess the qualities of a theory endowed with a field that eats up vacuum energy and makes up for cosmic acceleration. In this final subsection, we answer the important question: Can the self-tuning GCCG (Eqs. \eqref{eq:k_gccg_st} and \eqref{eq:g_gccg_st}) screen an \textit{arbitrarily} large bare cosmological constant? This is relevant to whether a theory can self-tune the particle vacuum's colossal energy density, $\rho_\Lambda$, to produce the observed cosmic acceleration. 

The answer is \textit{no}. To get to this conclusion, we rely strongly on the gradient stability constraint (Eq. \eqref{eq:ngc}), identifying $\rho_\Lambda$ and $h^2$ with the energy scales of the particle vacuum and observed cosmic acceleration, respectively. Clearly, when $\rho_\Lambda \gg h^2$, the self-tuning GCCG parameter $n$ approaches vanishingly small negative values, e.g., $n \rightarrow 0^-$ as $\rho_\Lambda / \left( 3 h^2 \right) \rightarrow \infty$. Even with the more relaxed input of $\rho_\Lambda / \left( 3 h^2 \right) \sim O \left( 50 \right)$ the gradient condition will constrain $n$ to be at most of order $| n | \sim O\left(-50\right)$, an admittedly unfair precision to expect in any cosmological surveys in the forthcoming years. Taking $n \rightarrow 0^-$ in Eqs. \eqref{eq:k_gccg_st} and \eqref{eq:g_gccg_st}, we find that the self-tuning theory reduces to the trivial KGB (a.k.a. a \textit{finely-tuned} cosmological constant) which naively renormalizes bare vacuum energy.

Notably, a gradient instability is not as serious compared with the existence of a ghost in vacuum. The absence of a gradient instability is nonetheless a respectable theoretical prior in constraining dark energy theories and is a reasonable basis for tightening the expectations on a self-tuning alternative gravity theory.

\section{Clash of titans: Large bare vacuum energy vs Self-tuning dark energy}
\label{sec:cc_vs_gradient}

In Sections \ref{sec:self_tuned_kgb} and \ref{sec:stgccg}, we have shown that self-tuning in shift symmetric KGB is indeed possible. However, in Section \ref{subsec:self_tuning_gccg_vs_large_cc}, we have also shown that shielding cosmic acceleration from an arbitrarily large vacuum energy, such as the one coming in from quantum fluctuations of the particle vacuum, can put a self-tuning theory in tension with the gradient stability constraint, essentially ruling out the GCCG theory as a self-tuning dark energy. This prompts the important question: Can cosmic acceleration be shielded from an \textit{arbitrarily} large bare cosmological constant by a self-tuning dark energy field in tadpole-free shift symmetric KGB?

The answer is \textit{yes} and we prove it with a concrete example. Unfortunately, we are unable to fully characterize this special set of self-tuning theories and so leave this open for future investigation.

Consider the self-tuning KGB given by the potentials
\begin{eqnarray}
\label{eq:k_exp_kgb} K \left( \phi, X \right) &=& c \exp \left( X / d \right) \\
\label{eq:g_exp_kgb} G \left( \phi, X \right) &=& - \dfrac{ c }{3h} \sqrt{ \dfrac{\pi}{2d} } \text{erfi} \left( \sqrt{\dfrac{X}{d}} \right)
\end{eqnarray}
where $c$ and $d$ are constants and $\text{erfi} \left( z \right) = \text{erf} \left( i z \right) / i$ is the imaginary error function. It is easy to show this satisfies Eq. \eqref{eq:b1}. In this theory, the Hamiltonian constraint on the de Sitter vacuum becomes
\begin{equation}
\label{eq:frd_eq_exp_theory}
3h^2 = \rho_\Lambda - c \exp \left( X / d \right)
\end{equation}
and the gradient stability condition can be broken down to
\begin{equation}
\label{eq:gradient_1_exp_theory}
d \left( 3 h^2 d + c X \exp \left( X/d \right) \right) > 0
\end{equation}
and
\begin{equation}
\label{eq:gradient_2_exp_theory}
c d \exp \left( X / d \right) < 0 .
\end{equation}
Looking at Eq. \eqref{eq:frd_eq_exp_theory}, we can see that $c > 0$ in order to screen $\rho_\Lambda \gg 3 h^2$. Eq. \eqref{eq:gradient_2_exp_theory} then implies that $d < 0$ and therefore Eq. \eqref{eq:gradient_1_exp_theory} simplifies to
\begin{equation}
\label{eq:gradient_1_exp_theory_2}
 -3 h^2 |d| + c X \exp \left( - X / | d | \right) < 0 .
\end{equation}
To show that the theory allows the self-tuning of $\rho_\Lambda \gg 3 h^2$, we nondimensionalize the variables as $\rho_\Lambda \rightarrow r \left( 3 h^2 \right)$, $X \rightarrow \chi \left( 3 h^2 \right)$, $c \rightarrow \gamma \left( 3 h^2 \right)$, and $d \rightarrow \Gamma  \left( 3 h^2 \right)$. Using the Hamiltonian constraint and defining the variable $\zeta = \exp \left( \chi / \Gamma \right)$, we can then express Eq. \eqref{eq:gradient_1_exp_theory_2} as
\begin{equation}
1 + \left( r - 1 \right) \ln \zeta > 0
\end{equation}
which is obviously satisfied for any $r = \rho_\Lambda / \left( 3 h^2 \right) \gg 1$ and $0 < \zeta < 1$. This exponential KGB (Eqs. \eqref{eq:k_exp_kgb} and \eqref{eq:g_exp_kgb}) can therefore tune away an arbitrarily large bare vacuum energy, affirming the positive answer to the important question posed in this section.

\begin{figure}[h!]
\center
	\subfigure[ ]{
		\includegraphics[width = 0.48 \textwidth]{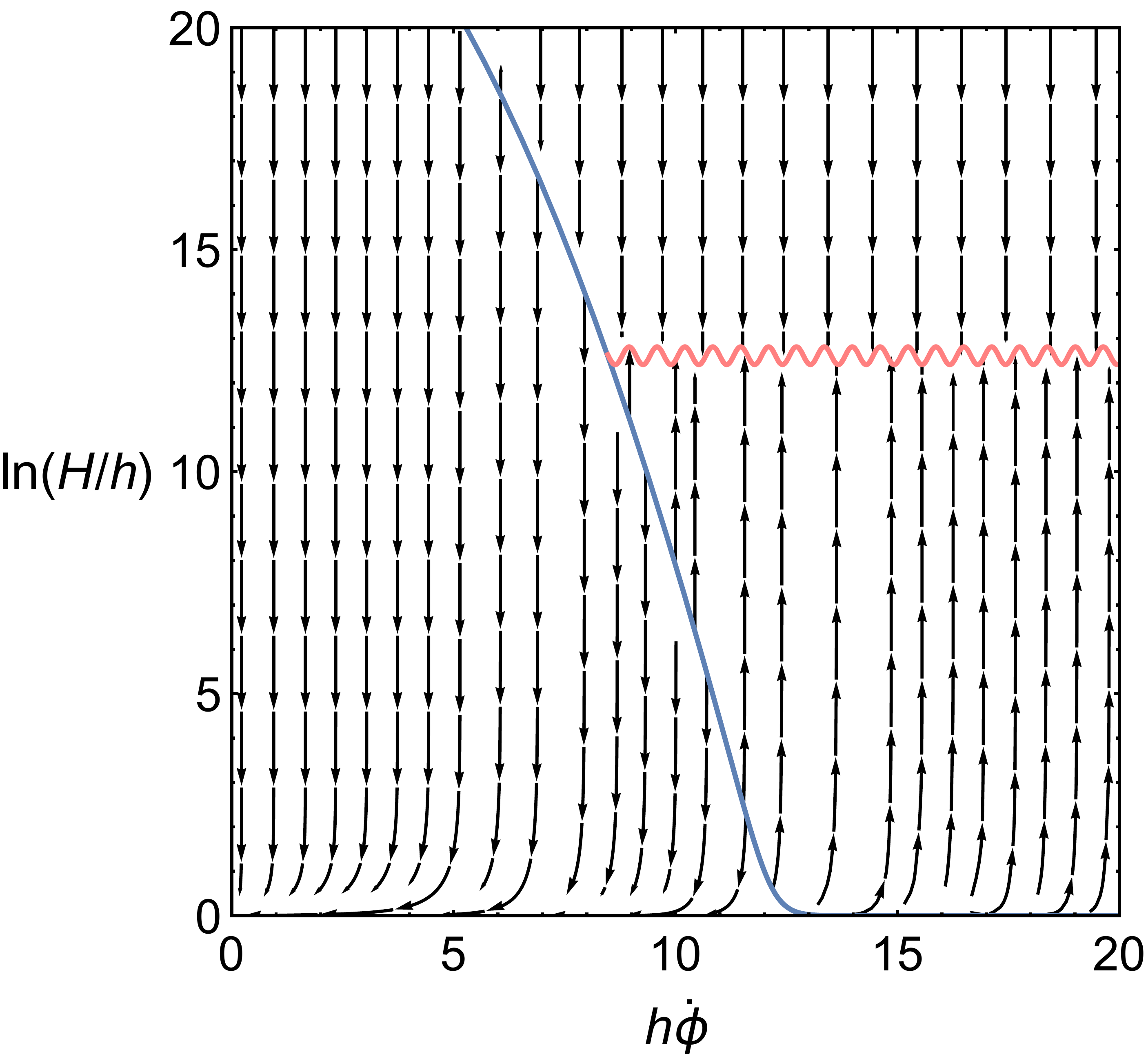}
		}
	\subfigure[ ]{
		\includegraphics[width = 0.48 \textwidth]{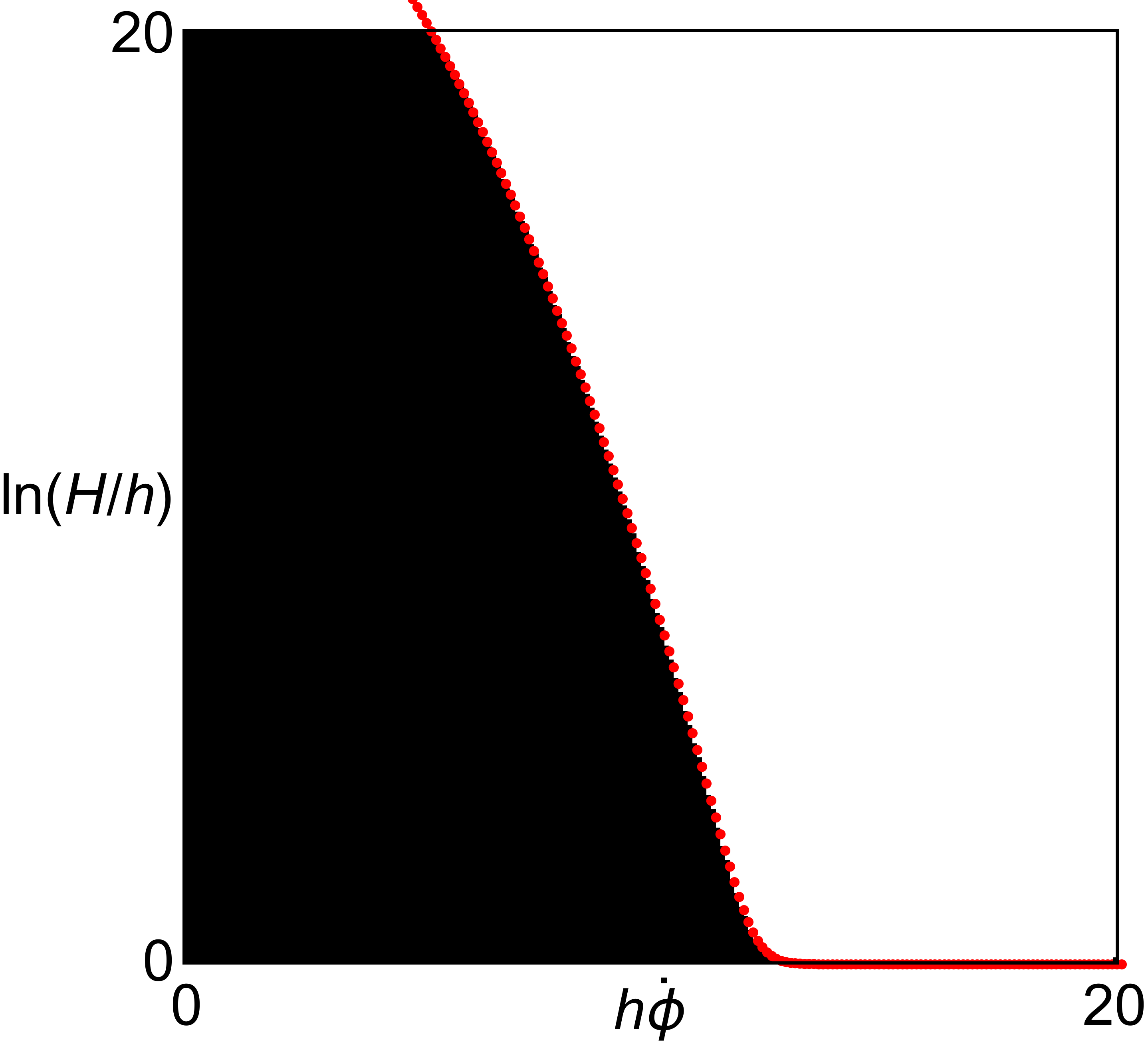}
		}
\caption{$\left[ \gamma = 9 \times 10^{10} - \left(1/2\right), \Gamma = -1 , \lambda = 9 \times 10^{10}, w = 0 \right]$ (a) Phase portrait of the vector field $(x', y')$ given by Eqs. \eqref{eq:dx_ekgb} and \eqref{eq:dy_ekgb}. Blue solid line is the critical curve $q \left( x, y \right) = 0$ and the pink wavy line marks $y^2 = \lambda$. (b) $200 \times 200$ pixels-plot of phase space obtained by numerically integrating Eqs. \eqref{eq:dx_ekgb} and \eqref{eq:dy_ekgb} and characterizing the late-time states. Red markers show the critical curve $q \left( x , y \right) = 0$ and black (white) region shows the section of phase space which falls to the healthy $y = 1$ (unhealthy $y^2 = \lambda$) asymptotic state.}
\label{fig:ds_exp_kgb}
\end{figure}

The same features found in the previous section for the GCCG theory also arise in the exponential KGB. We simply demonstrate the case with a single non-$\Lambda$ cosmic fluid. In this case, we can write down a two-dimensional dynamical system
\begin{equation}
\label{eq:dx_ekgb}
x' = \dfrac{3 x}{q\left(\tau\right)} \left( \gamma  e^{\frac{x^2}{6 \Gamma }} \left(3 \Gamma  (w+1)+w x^2 (y-1)\right)+3 \Gamma  (y((w-1) y+2)-\lambda  (w+1)) \right)
\end{equation}
and
\begin{equation}
\label{eq:dy_ekgb}
\begin{split}
y' = \dfrac{ y - 1 }{q\left(\tau\right)} \bigg( & 3 \Gamma  (w+1) \left(3 \Gamma +x^2\right) \left(\lambda -y^2\right) \\
& -\gamma  e^{\frac{x^2}{6 \Gamma }} \left(9 \Gamma ^2 (w+1)+3 \Gamma  x^2 ((w-1) y+1)+w x^4 (y-1)\right) \bigg)
\end{split}
\end{equation}
where
\begin{equation}
q \left( x, y \right) = 2 \Gamma  (y-1) \left(3 \Gamma +x^2\right)-\gamma  \Gamma  x^2 e^{\frac{x^2}{6 \Gamma }}
\end{equation}
and
\begin{eqnarray}
\gamma &=& c / 3 h^2 \\
\Gamma &=& d / 3 h^2 \\
\tau &=& h t \\
x \left( \tau \right) &=& h \dot{ \phi } \left( \tau/ h \right) \\
y \left( \tau \right) &=& H \left( \tau/ h \right) / h \\
\lambda &=& \rho_\Lambda / 3h^2 .
\end{eqnarray}
The stable fixed point corresponding to the healthy de Sitter vacuum can be found at $\left( x, y \right) = \left( \sqrt{ 6 \Gamma \ln \left( \left( \lambda - 1 \right) / \gamma \right) }, 1 \right)$, requiring $\lambda - 1 < \gamma$ since the gradient stability constraint demands $\Gamma < 0$. The other de Sitter state, $y^2 = \lambda$, can similarly be found but with $x \rightarrow \infty$. It can be confirmed that the gradient stability constraint is satisfied for parameters $\{ \gamma = 9 \times 10^{10} - \left(1/2\right), \Gamma = -1 , \lambda = 9 \times 10^{10}, w = 0 \}$. The corresponding phase portrait and basin plots are shown in figure \ref{fig:ds_exp_kgb}.
Indeed, the phase portrait (figure \ref{fig:ds_exp_kgb}(a)) confirms the existence of the two attractors, the self-tuning vacuum $y = 1$ (or $\ln \left( H / h \right) = 0$) and the unhealthy late-time attractor $y^2 = \lambda$. This is supported by the $200\times200$ pixels-basin plot (figure \ref{fig:ds_exp_kgb}(b)) obtained by integrating the dynamical system itself (Eqs. \eqref{eq:dx_ekgb} and \eqref{eq:dy_ekgb}) and classifying the phase space according to whether the late-time state of the numerical solution falls to $y = 1$ (shaded) or $y^2 = \lambda$ (unshaded). The viable phase space is the shaded region of figure \ref{fig:ds_exp_kgb}(b). The numerical value of $\lambda$ chosen for this simulation can be increased further by orders of magnitude, $\lambda = \rho_\Lambda / \left( 3h^2 \right) > 10^{10}$, without spoiling the gradient stability constraint as long as computational resources are available.

\section{Self-tuning in KGB: In and out of shift symmetry}
\label{sec:tempered_trivial_scalar}

The cosmological constant problem is arguably the biggest mystery in theoretical physics and for this reason it is desirable to identify alternative gravity theories with self-tuning fields. KGB belongs to this important subset and its well-tempering class has been the subject of Refs. \cite{st_horndeski_cubic_appleby, st_horndeski_cosmology_emond2018, st_horndeski_variations_appleby, st_fugue_appleby}.
\begin{figure}[h!]
\center
\includegraphics[width = 0.7 \textwidth]{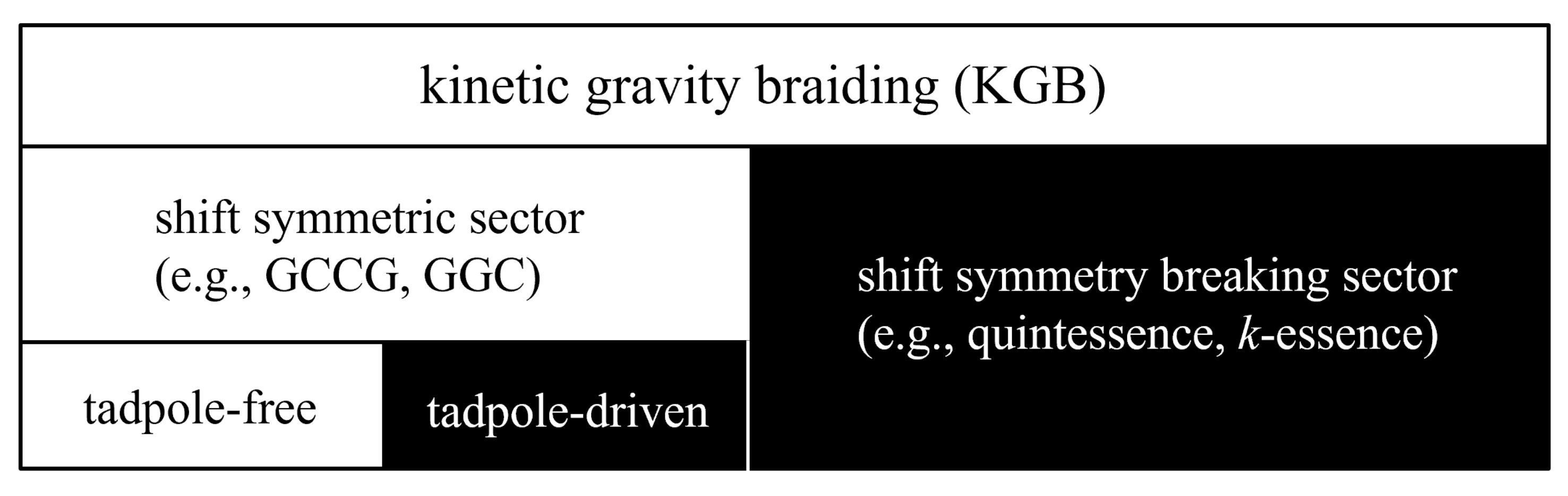}
\caption{Self-tuning mechanisms in kinetic gravity braiding: \textit{well-tempering} applies to the shaded regions (tadpole-driven, shift symmetry breaking sector) and \textit{trivial scalar} self-tuning applies to all KGB. GCCG and GGC stand for generalized cubic covariant Galileon and Galileon ghost condensate, respectively.}
\label{fig:kgb_self_tuning}
\end{figure}
This paper, on the other hand, fully exposes the self-tuning branch of the tadpole-free shift symmetric KGB and is complimentary to the existing literature on the subject (figure \ref{fig:kgb_self_tuning}). We have shown using analytical results and various numerical experiments that the self-tuning vacuum of tadpole-free theory is an attractor and can be free from ghost and gradient instabilities.

It should be mentioned that it is possible for theories in the shaded region of figure \ref{fig:kgb_self_tuning} to accommodate both tempered and trivial scalar de Sitter states \cite{st_horndeski_cubic_appleby}. See example in Appendix B of Ref. \cite{st_horndeski_cubic_appleby}. The reader interested in the well-tempering subclass of KGB beyond the scope of the discussion of Section \ref{subsec:well_tempered_kgb} is encouraged to consult with Refs. \cite{st_horndeski_cubic_appleby, st_horndeski_cosmology_emond2018, st_horndeski_variations_appleby, st_fugue_appleby}.

Tadpole-free shift symmetric KGB remains to be a phenomenologically viable sector of KGB \cite{peirone2019cosmological, gccg_frusciante}. It is worth pointing out that this sector can accommodate nearly arbitrary cosmological dynamics and contains hairy black holes \cite{bernardo2019hair, stealth_bernardo, self_tuning_bh_emond, bernardo2019tailoring, designing_horndeski_arjona}. A highlight of this paper is the no-tempering theorem of Section \ref{subsec:well_tempered_kgb} which is also illustrated in figure \ref{fig:kgb_self_tuning}. Outside of tadpole-free shift symmetric KGB, both well-tempering and trivial scalar self-tuning can be achieved. 

The main issue with the unavailability of well-tempering in tadpole-free shift symmetric KGB is that the self-tuning field screens all cosmic fluids instead of only vacuum energy (Sec. \ref{subsec:self_tuned_kgb}). It is an open question if this is resolvable within the theory alone, e.g., using dark sector interactions, or if it is inevitable to resort to well-tempering through the existence of the tadpole or by going beyond shift symmetry.

\section{Outlook}
\label{sec:outlook}

We have shown that self-tuning can be achieved in tadpole-free shift symmetric KGB, despite the absence of well-tempering, and that the resulting theory possesses some pleasant features, e.g., the inevitability of cosmic acceleration and avoidance of ghosts. However, there remains open issues, the big one being that the implemented self-tuning mechanism blocks all energy, instead of only vacuum energy, resulting in a questionable dark energy theory. The classification of tadpole-free shift symmetric KGB that can tune away an arbitrarily large bare vacuum energy is also left open for further investigation.

The consequences of self-tuning on perturbations, e.g., on the growth rate and eventual collapse of matter overdensities, also remains to be worked out. The modified equation for the linear density contrast is straightforward to obtain in terms of a single phenomenological function in the quasistatic regime in KGB. However, it must be recognized that a theory parametrized by a single free potential can still be considered to be quite large and the validity of the quasistatic approximation for the self-tuning corners of KGB also needs to be tested \cite{st_horndeski_lss_peirone, cosmo_horndeski_frusciante, cosmo_pert_lagos, qsa_horndeski_pace}. The details of this perturbative analysis is left for future work.

There are several other places where self-tuning can be further investigated. To begin, KGB was introduced almost simultaneously in the context of dark energy and primordial inflation. Primordial inflation is also an active field where a self-tuning mechanism may be potentially realizable. It remains to be explored whether self-tuning can be achieved in scalar-tensor theories outside of the KGB class and with luminal gravitational wave propagation \cite{dhost_langlois, dhost_achour, dark_energy_dhost_crisostomi}. There are also very promising approaches to the cosmological constant problem that are outside of classical well-tempering and trivial scalar self-tuning (see Refs. \cite{cc_lombriser_1, cc_lombriser_2, self_tuning_blanco_1, self_tuning_blanco_2}).

Lastly, it is worth looking for the crossroads between self-tuning and other desirable features of a cosmological theory such as the existence of an Einstein gravity limit and scaling solutions. A viable dark energy theory must also be accompanied by a screening mechanism to guarantee compatibility with Solar system tests. The strong gravity regime of a self-tuning theory can also be expected to be rich with observational signatures and should be given due attention in light of active and forthcoming gravitational wave surveys.

\acknowledgments
The author is grateful to Stephen Appleby for many insightful remarks on an earlier draft of this paper and to Sean Fortuna and Joshua Bautista for helpful suggestions in numerical execution.







\bibliographystyle{JHEP}
\providecommand{\href}[2]{#2}\begingroup\raggedright\endgroup

\end{document}